\documentclass[]{elsarticle}

\journal{Journal of Atmospheric and Solar-Terrestrial Physics}

\usepackage{lineno, hyperref}
\modulolinenumbers[5]

\usepackage{graphicx}        
\usepackage{amssymb}         
\usepackage{color}           

\newcommand{\sech}{ \mathrm{sech}}

\newcommand{\aap}{    {\it Astron. Astrophys.}}

\newcommand{\apj}{    {\it Astrophys. J.}}
\newcommand{\apjl}{   {\it Astrophys. J. Lett.}}

\newcommand{\pasj}{   {\it Pub. Astron. Soc. Japan}}

\newcommand{\solphys}{{\it Solar Phys.}}
 
\newcommand{\ssr}{    {\it Space Sci. Rev.}} 
\newcommand{\pra}{    {\it Phys. Rev. A}}
\newcommand{\zap}{    {\it Zeitschrift f\"ur Astrophysik}}
\newcommand{\caa}{    {\it Chin. Astron. Astrophys.}}
\chardef\us=`\_

\biboptions{authoryear}

\begin{document}

\begin{frontmatter}

\title{{Finite amplitude transverse oscillations of a magnetic rope}}

\author[aff1]{Dmitrii Y. Kolotkov\corref{mycorrespondingauthor}}
\cortext[mycorrespondingauthor]{Corresponding author}
\ead{D.Kolotkov@warwick.ac.uk}

\author[aff1,aff2]{Giuseppe Nistic\`o}

\author[aff1]{George Rowlands}

\author[aff1,aff3]{Valery M. Nakariakov}

\address[aff1]{Centre for Fusion, Space and Astrophysics, Department of Physics, University of Warwick, CV4 7AL, UK}
\address[aff2]{Now at Georg-August-Universit\"at G\"ottingen, Germany}
\address[aff3]{St. Petersburg Branch, Special Astrophysical Observatory, Russian Academy of Sciences, 196140, St. Petersburg, Russia}

\begin{abstract}
{The effects of finite amplitudes on the transverse oscillations of a quiescent prominence represented by a magnetic rope are investigated in terms of the model proposed by \citet{Kolotkov2016}. We consider a weakly nonlinear case governed by a quadratic nonlinearity, and also analyse the fully nonlinear equations of motion.}
{We treat the prominence as a massive line current located above the photosphere and interacting with the magnetised dipped environment via the Lorentz force. In this concept the magnetic dip is produced by two external current sources located at the photosphere.}
{{Finite} amplitude horizontal and vertical oscillations are found to be strongly coupled between each other. The coupling is more efficient for larger amplitudes and smaller attack angles between the direction of the driver and the horizontal axis. Spatial structure of oscillations is represented by Lissajous-like curves with the limit cycle of a hourglass shape, appearing in the resonant case, when the frequency of the vertical mode is twice the horizontal mode frequency. A metastable equilibrium of the prominence is revealed, which is stable for small amplitude displacements, and becomes horizontally unstable, when the amplitude exceeds a threshold value. The maximum oscillation amplitudes are also analytically derived and analysed. Typical oscillation periods are determined by the oscillation amplitude, prominence current, its mass and position above the photosphere, and the parameters of the magnetic dip.}
{The main new effects of the {finite} amplitude are the coupling of the horizontally and vertically polarised transverse oscillations (i.e. the lack of a simple, elliptically polarised regime) {and the presence of metastable equilibria of prominences.}}
\end{abstract}

\begin{keyword}
quiescent prominences \sep filaments \sep nonlinear oscillations \sep magnetic field 
\end{keyword}

\end{frontmatter}


\section{Introduction}
Solar prominences are the condensations of plasma at temperatures of about $10^4$~K (typical for the chromosphere) floating in the much hotter solar corona (with temperatures typically greater than $10^6$ K) \citep[see e.g.][for a comprehensive review]{Parenti2014}.
The main questions related to prominences concern the physical mechanisms involved in their formation and evolution. Indeed, prominences can be generally distinguished in two categories: quiescent prominences, which are observed floating in the low solar corona with time scales ranging from hours to several days before to slowly fade out or dissolve; and erupting prominences, which become unstable in the presence of particular physical conditions. As a consequence of the prominence eruption, a coronal mass ejection (CME) could be formed and expelled from the solar corona.
The loss of equilibrium can be caused by various reasons: eruptions can be triggered by a nearby flare \citep{2015ApJ...811....5P}, or in response to an emerging magnetic flux or variation of the local magnetic helicity \citep{Yeates2009}, or maybe due to the action of MHD waves, as observed for some events before the eruption onset \citep[see e.g. the discussion in][]{2014ApJ...786..151S}. Quiescent prominences are also very dynamic, being a subject to MHD oscillations \citep{2012LRSP....9....2A}, such as transverse oscillations, for example triggered by a global coronal wave \citep[e.g.][]{Hershaw2011, Asai2012}, and longitudinal oscillations \citep[e.g.][]{2007A&A...471..295V, 2012A&A...542A..52Z, Luna2014}.
In turn, based on the direction of the filament main axis displacements, transverse oscillations can have horizontal \citep[e.g.][]{1969SoPh....6...72K, Hershaw2011, 2012ApJ...753...53S}, or vertical polarisations \citep[e.g.][]{1966ZA.....63...78H, 2002PASJ...54..481E, 2004ApJ...608.1124O, 2014ApJ...797L..22K, 2016ARep...60..287M}. {Furthermore, quiescent prominence threads are also observed to experience more complicated, chaotic, spatial dynamics during large amplitude oscillations \citep[see e.g.][]{2008ApJ...685..629G, 2017ApJ...836..178T}.}
Complex behaviour of plasma in prominences can be also described in terms of turbulent processes \citep{Berger2010,Leonardis2012}. Such evidences may be strongly affected by thermodynamic processes acting in prominences, which can also influence the evolution of slow MHD waves \citep[][]{Kumar2016, Ballester2016}. In addition, Kelvin--Helmholtz instability may take place during oscillations of prominences, sustaining damping and plasma heating \citep[][]{Antolin2014,Terradas2016}. Also, the presence of continuous transverse oscillations in prominences \citep{2013ApJ...779L..16H} may also be referred to as a self-oscillatory process caused by the interaction of plasma nonuniformities with a quasi-steady flow \citep{2016A&A...591L...5N}. 

The equilibrium of prominences is thought to be of a magnetic origin with the Lorentz force counteracting the gravity. In turn, gradient pressure forces can provide an additional support. Considering this basic idea, the following two-dimensional (2D) models of the prominence equilibrium are the most popular: the Kippenhahn--Schl\"uter \citep[KS,][]{KS1957} and the Kuperus--Raadu models \citep[KR,][]{KR1974}. The KS model considers the prominence as a plasma slab embedded in the straight magnetic field lines with a dip created by some external sources (e.g. photospheric currents). The magnetic dip outlines a region of magnetic polarity inversion, which justifies a general empirical evidence that prominences lie along the polarity inversion line (also called a neutral line) of large extended bipolar regions \citep[e.g.][]{Bosman2012}. 
In the KR model the prominence is assumed to be a straight current-carrying horizontal wire located at some height above the conductive photosphere. The support against the gravity is provided by an upward magnetic force acting on the prominence and caused by a virtual \lq\lq mirror\rq\rq\ current, which is located below the photosphere and strictly symmetrical to the prominence. Interestingly, the magnetic topology associated with the KR model resembles that of a  coronal cavity, that is a large quasi-circular structure observed off limb in the extreme ultraviolet (EUV) band, and containing a prominence in its interior \citep{Habbal2010,Gibson2010}.

{In the last decades, starting from these two seminal works of KS and KR, a number of studies of 2.5D and full 3D models of prominences have been carried out, taking into account such observational aspects as the presence of a current-aligned magnetic field component, magnetic chirality, \lq\lq barbs\rq\rq\ or \lq\lq feet\rq\rq\ connecting the prominence to the photosphere, H$\alpha$ fibrils, flows, and their association to CMEs in case of eruptions. In this context, modelling of prominences supported in twisted flux tubes (magnetic flux ropes) by linear force-free field was undertaken by \citet{1998A&A...329.1125A} and \citet{1998A&A...335..309A}, addressing the natural presence of lateral feet and fibrils. A further approach is to consider extrapolations from photospheric magnetic field data, and compare measurements of prominence locations with the local dips in the resulting coronal magnetic field configurations \citep{2003A&A...402..769A, 2012ApJ...757..168S}. \citet{2011A&A...532A..93B} studied magneto-hydrostatic (MHS) equilibria for prominences by reducing the MHS equations to an extended Grad--Shafranov equation, and then numerically investigated the spectra of the oscillating structure.  A relaxation process is another approach to study the effect of support against the gravity by the magnetic field, where the cold and dense prominence plasma is injected into an initially unperturbed background, and the subsequent evolution is studied numerically. \citet{2013ApJ...766..126H} studied equilibria for two distinct magnetic field structures of an inverse polarity: a simple o-point configuration, and a more complex one with an x-point. In the former case, the magnetic tension of the field lines compressed at the base of the prominence and stretched at its top is able to sustain prominences, while in the latter case a convergence to a prominence equilibrium is not always guaranteed. \citet{2013ApJ...778...49T} investigated properties of MHD waves in normal polarity prominences embedded in coronal arcades in terms of the relaxation model too. Stable vertical fast and longitudinal slow MHD oscillations were found. \citet{2012ApJ...757...98L} and \citet{2016SoPh..291..429K} also considered prominences of a normal configuration, residing in a dip formed by curved magnetic field lines. The effects of the magnetic field geometry on longitudinal oscillations in prominences were addressed.}

{Despite their exceptional importance,} the KS and KR models separately are not able to provide an exhaustive picture on the transverse oscillations observed in prominences. For example, the KR model alone allows only for vertically polarised oscillations, while in the pure KS model horizontally polarised oscillations cannot coexist with the vertically polarised ones since the system becomes unstable \citep{Oord1998}.
A synthesis of these two models, that is a prominence embedded in a magnetic field dip generated by two photospheric currents, accounting also for the effects of the prominence current interaction with the conducting photosphere (via the inclusion of the mirror current effect), has been recently developed in \citet[][KNN16]{Kolotkov2016}. The prominence has been modelled as a line current located above the photosphere at a given height, thus being subject to the gravity and Lorentz forces, which are attributed to the interaction between the photospheric and prominence currents. Such a magnetostatic model, despite its simplicity, provides straightforward results on the prominence dynamics. In KNN16, horizontally and vertically polarised transverse oscillations have been analysed in the linear regime, the equations of motion analytically derived, and dependence of the oscillation properties (e.g. the period) upon the parameters of the system (e.g. the currents in the prominence and at the photosphere) has been determined. In addition, investigation of the mechanical stability of the system shows that the prominence can be stable simultaneously in both horizontal and vertical directions for a certain range of parameters. 

In this work, we study oscillations of finite amplitude in terms of the KNN16 model, addressing two main issues: determining the domain of the applicability of the linear approximation derived in KNN16, and responding to the observational detection of finite amplitude oscillations in prominences \citep[e.g.][]{Tripathi2009}.  We show that the equations of motion in the vertical and horizontal directions are nonlinearly coupled with each other, in contrast to the linear regime where the motions are essentially independent of each other. Therefore, the presence of nonlinear terms in the governing equations makes the dynamics of the system more various and rich. The paper is structured as follows: in Sect. 2 we present the model and the governing equations; in Sect. 3 we provide an analytical treatment of the equations of motion along the vertical and horizontal directions in the presence of a weak nonlinearity, in Sect. 4 we present an analysis of the oscillation amplitudes and periods by the consideration of a total energy of the system. Finally, discussion and conclusions are provided in Sect. 5.

\section{Model and governing equations}
\begin{figure}
	\begin{center}
		\includegraphics[width=8cm]{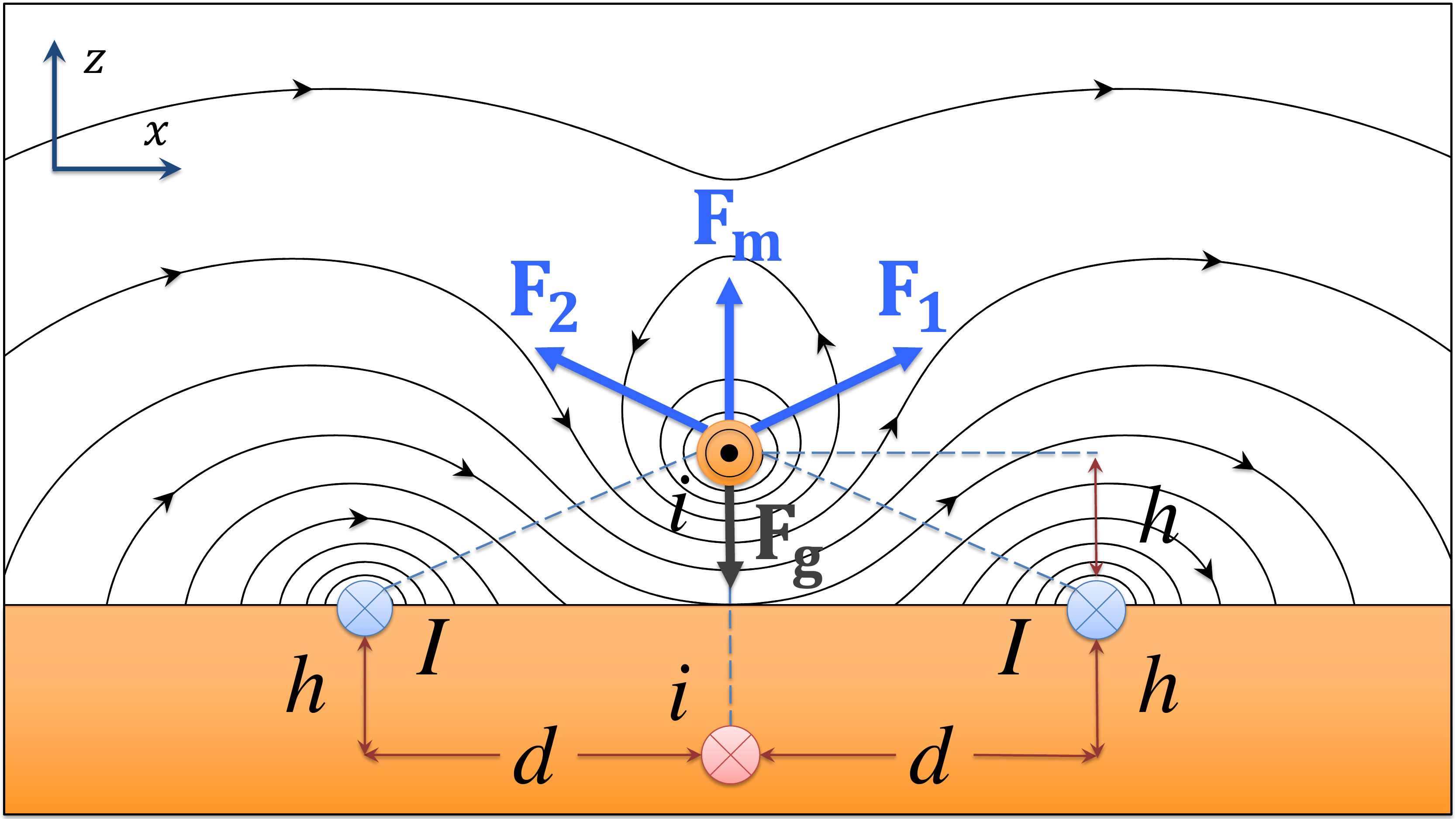}
	\end{center}
	\caption{{A massive prominence (indicated by the yellow blob) with a line current $i$, located in a magnetic dip at the height $h$ above the photosphere. The dip is formed by two external current sources $I$ separated by the distance $2d$ at the photosphere. The mirror current $i$ (the red blob) is located strictly below the prominence. The magnetic field lines produced by these four currents are shown for $h=0.5\,d$ and $i=0.5\,I$ \citep[similar to Fig.~1 in][]{Kolotkov2016}.}}
	\label{sketch}
\end{figure}

Consider a prominence as a horizontal line current $i$, located at the height $h$ above the plane photosphere in a magnetic dip produced by two spatially separated photospheric line currents of the same strength $I$ parallel to the prominence current, with $d$ being the half-distance between them (see Fig.~\ref{sketch}, where the origin of the coordinate system coincides with the centre of the equilibrium current in the unperturbed prominence). {The magnetic configuration shown in Fig.~\ref{sketch} corresponds to a normal polarity prominence, i.e. the polarity of the magnetic field lines threading the prominence material coincides with that of the underlying photospheric field \citep[cf. Fig.~2 in][]{2002ApJ...564L..53L}. Although prominences of this type constitute about 10\% to 25\% of the observed prominences \citep[see e.g.][]{1984A&A...131...33L, 1994SoPh..154..231B, Parenti2014, 2017ApJ...835...94O}, the flux ropes with a normal configuration are usually observed in the vicinity of active regions \citep[see e.g.][]{2008ApJ...673L.215O, 2010ApJ...714..343G, 2012A&A...539A.131K, 2014A&A...561A..98S}, and can be responsible for fast CMEs \citep[][]{2002ApJ...564L..53L}.}
The horizontal equilibrium of the prominence in such a magnetic system is provided automatically because of the horizontal symmetry of the model, while the vertical equilibrium {is determined by the balance of the gravity force $\bf F_g$ and three Lorentz forces $\bf F_1$, $\bf F_2$, and $\bf F_m$ acting on the prominence from the external photospheric and mirror currents, respectively. In the projection onto the $z$-axis, the vertical equilibrium condition is}
\begin{equation}\label{equilibrium}
\frac{2k_1h}{d^2+h^2}+\frac{k_2}{2h}=\mathcal{R} g,
\end{equation}
with $k_1=\mu_0 I i/2 \pi$, $k_2=\mu_0 i^2/2 \pi$. In (\ref{equilibrium}) $\mathcal{R} g$ denotes the gravity force per unit length assumed to be constant in the model, with the linear mass density $\mathcal{R}$ obtained as the volume mass density of the prominence, multiplied by its cross-sectional area. In condition (\ref{equilibrium}) and in the following analysis we consider the forces normalised per unit length in the direction parallel to the currents.

Dynamics of such a prominence, perturbed by an oblique displacement with the corresponding $x$ (horizontal) and $z$ (vertical) components, is governed by the following set of equations:
\begin{equation}\label{Fx}
\mathcal{R} \frac{d^2\,x}{d\,t^2}=F_x,
\end{equation}
\begin{equation}\label{Fz}
\mathcal{R} \frac{d^2\,z}{d\,t^2}=F_z,
\end{equation}
where
$$F_x=\frac{2k_1x[(h+z)^2+x^2-d^2]}{(d^2-x^2)^2+2(d^2+x^2)(h+z)^2+(h+z)^4}\nonumber,$$
$$F_z=\frac{2k_1(h+z)[d^2+x^2+(h+z)^2]}{(d^2-x^2)^2+2(d^2+x^2)(h+z)^2+(h+z)^4}+\frac{k_2}{2h+z}-\mathcal{R} g.\nonumber$$
{We have to point out that in this study we do not take into account any dissipative effects. This is justified by the need to develop the analytical formalism that can provide important insights in the main features of the oscillatory properties of prominences. Moreover, the limited duration of the detected oscillations does not necessarily indicate the presence of the oscillation damping, and may be caused by the change of the observational conditions. Thus, it is not clear which dissipative processes have to be included in the model. In any case, the formalism to be developed in this work is applicable to the modelling of the initial phase of transverse oscillations of prominences, before some dissipative processes cause the oscillation damping.}

\begin{figure}
	\begin{center}
		\includegraphics[width=7cm]{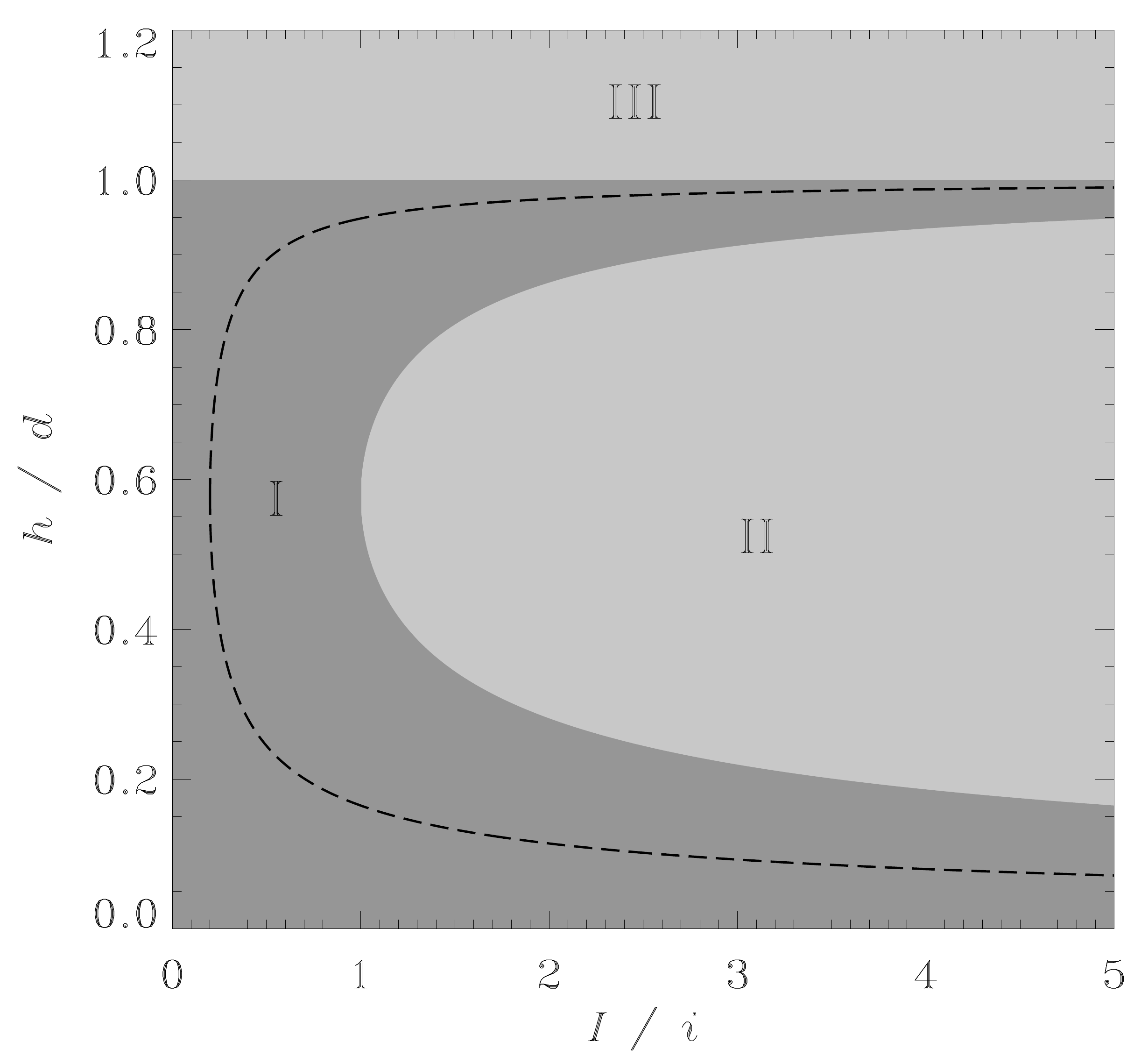}
	\end{center}
	\caption{Diagram showing the regions of parameters, where the prominence is stable in both vertical and horizontal directions simultaneously (region I); unstable in the vertical direction (region II); and unstable in the horizontal direction (region III). The regions were obtained in the linear theory developed in \citet[][KNN16]{Kolotkov2016}. The dashed line shows the nonlinear resonance condition between vertical and horizontal oscillatory modes, determined by Eq.~(\ref{res}).}
	\label{regions}
\end{figure}

Linear oscillatory solutions of Eqs.~(\ref{Fx}) and (\ref{Fz}) have been recently analysed in KNN16, treating the displacements $x$ and $z$ to be small and, hence, using the first order Taylor expansions of Eqs.~(\ref{Fx}) and (\ref{Fz}). In this linear regime the regions of parameters, corresponding to fully stable prominence oscillations (region I, Fig.~\ref{regions}), vertical instability (region II, Fig.~\ref{regions}), and horizontal instability (region III, Fig.~\ref{regions}), were revealed. Moreover, the linear vertical and horizontal modes were found to be essentially decoupled, therefore, can be considered separately. However, in the case of {finite} amplitude oscillations the coupling between the vertically and horizontally polarised modes cannot be ignored and must be taken into account.

\section{Weakly nonlinear coupling and resonance of vertically and horizontally polarised oscillations}
\label{weak}

With the use of the Taylor expansion up to the second order of the displacements $x$ and $z$, Eqs.~(\ref{Fx}) and (\ref{Fz}) can be re-written as
\begin{equation}\label{Fx2}
\frac{d^2\,x}{d\,t^2}=\alpha x+\beta xz,
\end{equation}
\begin{equation}\label{Fz2}
\frac{d^2\,z}{d\,t^2}=\gamma z+\delta x^2+\sigma z^2,
\end{equation}
where
$$\alpha=\frac{2k_1(h^2-d^2)}{\mathcal{R} (d^2+h^2)^2},~~~~\beta=\frac{4k_1h(3d^2-h^2)}{\mathcal{R} (d^2+h^2)^3}\nonumber,$$
$$\gamma=\frac{2k_1(d^2-h^2)}{\mathcal{R} (d^2+h^2)^2}-\frac{k_2}{4\mathcal{R} h^2},~~~~\delta=\frac{2k_1h(3d^2-h^2)}{\mathcal{R} (d^2+h^2)^3}\nonumber,$$
$$\sigma=\frac{2k_1h(h^2-3d^2)}{\mathcal{R} (d^2+h^2)^3}+\frac{k_2}{8\mathcal{R} h^3}\nonumber.$$
In contrast to the first order expansion of Eqs.~(\ref{Fx}) and (\ref{Fz}), considered in KNN16, Eqs.~(\ref{Fx2}) and (\ref{Fz2}) are coupled through the second order terms on the right-hand sides of (\ref{Fx2}) and (\ref{Fz2}).

{The set of coupled nonlinear equations (\ref{Fx2})--(\ref{Fz2}) represents a conservative system and for certain values of the parameters $\alpha$, $\beta$, $\gamma$, $\delta$, and $\sigma$,  was previously found to be integrable with the Hamiltonian of a H\'enon--Heiles form \citep[see e.g. {Eqs.~(3.1) in}][{the special case of $\beta =2$ and $\delta =1$}]{1982PhRvA..25.1257B}.
	In the present analysis we obtain general solutions of Eqs.~(\ref{Fx2})--(\ref{Fz2}), allowing for arbitrary values of those parameters, using the perturbation theory approach.}
Expressing the displacements $x$ and $z$ through a small parameter $\epsilon$ as $x\equiv\epsilon x$ and $z\equiv\epsilon z$ and expanding the new $x$ and $z$ with respect to $\epsilon$, $x=x_0+\epsilon x_1$ and $z=z_0+\epsilon z_1$, one can re-write Eqs.~(\ref{Fx2}) and (\ref{Fz2}) as
\begin{equation}\label{eps_expanded_x}
\frac{d^2\,x_0}{d\,t^2}+\epsilon \frac{d^2\,x_1}{d\,t^2}=\alpha x_0+\epsilon (\alpha x_1+\beta x_0z_0),
\end{equation}
\begin{equation}\label{eps_expanded_z}
\frac{d^2\,z_0}{d\,t^2}+\epsilon \frac{d^2\,z_1}{d\,t^2}=\gamma z_0+\epsilon (\gamma z_1+\delta x_0^2+\sigma z_0^2).
\end{equation}
The parameter $\epsilon$ demonstrates the smallness of the prominence displacements in comparison with the equilibrium geometrical parameters $d$ and $h$.
In such a representation of $x$ and $z$, the lowest order terms, $x_0$ and $z_0$, correspond to the decoupled harmonic oscillations of the prominence, while the higher-order components, $x_1$ and $z_1$, describe, in particular, the coupling between the horizontal and vertical modes. Indeed, collecting together the terms with the lowest order of the parameter $\epsilon$ in Eqs.~(\ref{eps_expanded_x}) and (\ref{eps_expanded_z}), one obtains
\begin{equation}\label{harmonicx}
\frac{d^2\,x_0}{d\,t^2}+\omega_1^2x_0=0,
\end{equation}
\begin{equation}\label{harmonicz}
\frac{d^2\,z_0}{d\,t^2}+\omega_2^2z_0=0,
\end{equation}
where $\omega_1^2=-\alpha$ and $\omega_2^2=-\gamma$. Behaviour of $\omega_1$ and $\omega_2$  and their dependence on the geometrical parameters of the model, $h$ and $d$, magnetic constants, $k_1$ and $k_2$, and the prominence mass density, $\mathcal{R}$, and the associated linear oscillations, has been investigated in detail in KNN16 model{, where the notations $\omega_1 = 2\pi/P_x$ and $\omega_2 = 2\pi/P_z$ were used, with $P_x$ and $P_z$ being the horizontal and vertical oscillation periods, respectively.} Eqs.~(\ref{harmonicx}) and (\ref{harmonicz}) have harmonic solutions written as
\begin{equation}\label{harmonicx2}
x_0(t)=A_1 \sin (\omega_1t+\phi_1),
\end{equation}
\begin{equation}\label{harmonicz2}
z_0(t)=A_2 \sin (\omega_2t+\phi_2),
\end{equation}
where $A_1$, $A_2$, $\phi_1$, and $\phi_2$ are the constants determined from the initial conditions.

Then we combine the terms of the first order of $\epsilon$ in Eqs.~(\ref{eps_expanded_x}) and (\ref{eps_expanded_z}). This gives
\begin{eqnarray}
\label{nonharmonicx}
\frac{d^2\,x_1}{d\,t^2}+\omega_1^2x_1=\frac{A_1A_2\,\beta}{2}\left\{\cos\left[(\omega_1-\omega_2)t+(\phi_1-\phi_2)\right]-\right.\\ \nonumber
\left.\cos\left[(\omega_1+\omega_2)t+(\phi_1+\phi_2)\right] \right\},
\end{eqnarray}
\begin{equation}\label{nonharmonicz}
\frac{d^2\,z_1}{d\,t^2}+\omega_2^2z_1=A_1^2\delta \sin^2(\omega_1t+\phi_1)+A_2^2\sigma \sin^2(\omega_2t+\phi_2),
\end{equation}
where the solutions for $x_0(t)$ and $z_0(t)$, given by Eqs.~(\ref{harmonicx2}) and (\ref{harmonicz2}), have been used. Solutions of Eqs.~(\ref{nonharmonicx})--(\ref{nonharmonicz}) can be written in a general form as
\begin{eqnarray}\label{nonharmonicx2}
x_1(t)=B_1\sin (\omega_1t+\psi_1)+\frac{A_1A_2\,\beta}{2}\left\{\frac{\cos[(\omega_1-\omega_2)t+(\phi_1-\phi_2)]}{\omega_2(2\omega_1-\omega_2)}+\right.\\ \nonumber
\left.\frac{\cos[(\omega_1+\omega_2)t+(\phi_1+\phi_2)]}{\omega_2(2\omega_1+\omega_2)} \right\},
\end{eqnarray}
\begin{eqnarray}\label{nonharmonicz2}
z_1(t)=B_2\sin(\omega_2t+\psi_2)+\frac{A_1^2\delta}{2}\frac{\cos[2(\omega_1t+\phi_1)]}{4\omega_1^2-\omega_2^2}+\nonumber\\ \sigma A_2^2\frac{\cos[2(\omega_2t+\phi_2)]}{6\omega_2^2}+\frac{\delta A_1^2+\sigma A_2^2}{\omega_2^2},
\end{eqnarray}
where $B_1$, $B_2$, $\psi_1$, and $\psi_2$ are the constants determined from the initial conditions.

Thus, combining solutions (\ref{harmonicx2})--(\ref{harmonicz2}) for $x_0$ and $z_0$ and (\ref{nonharmonicx2})--(\ref{nonharmonicz2}) for $x_1$ and $z_1$, and recalling that $x=x_0+\epsilon x_1$ and $z=z_0+\epsilon z_1$, the oscillatory solution of Eqs.~(\ref{Fx2})--(\ref{Fz2}) can be written as
\begin{eqnarray}\label{solx}
x(t)=C_1\sin (\omega_1t+\Theta_1)+\frac{C_1C_2\,\beta}{2}\left\{\frac{\cos[(\omega_1-\omega_2)t+(\Theta_1-\Theta_2)]}{\omega_2(2\omega_1-\omega_2)}+\right.\nonumber\\
\left.\frac{\cos[(\omega_1+\omega_2)t+(\Theta_1+\Theta_2)]}{\omega_2(2\omega_1+\omega_2)} \right\},
\end{eqnarray}
\begin{equation}\label{solz}
z(t)=C_2\sin(\omega_2t+\Theta_2)+\frac{\delta C_1^2}{2}\frac{\cos[2(\omega_1t+\Theta_1)]}{4\omega_1^2-\omega_2^2}+\nonumber\\\sigma C_2^2\frac{\cos[2(\omega_2t+\Theta_2)]}{6\omega_2^2},
\end{equation}
where $C_{1,2}\equiv [A_{1,2}^2+B_{1,2}^2+2A_{1,2}B_{1,2}\cos(\phi_{1,2}-\psi_{1,2})]^{1/2}$ and $\tan(\Theta_{1,2})=[A_{1,2}\sin(\phi_{1,2})+B_{1,2}\sin(\psi_{1,2})]/[A_{1,2}\cos(\phi_{1,2})+B_{1,2}\cos(\psi_{1,2})]$, with $\epsilon =1$. The use of $\epsilon =1$ in expressions (\ref{solx})--(\ref{solz}) does not contradict to the sense of generality as it was employed only for the quantification of the smallness of amplitudes of the higher order components ($B_1$, $B_2$, $A_1^2$, $A_2^2$, and $A_1A_2$) in comparison with the lowest harmonic amplitudes, $A_1$ and $A_2$.
The set of solutions (\ref{solx})--(\ref{solz}) describes the coupled horizontal and vertical oscillatory dynamics of the prominence and, importantly, implies a nonlinear resonance condition $2\omega_1=\omega_2$, appearing in both polarizations simultaneously. One can re-write this resonance condition in terms of the intrinsic physical parameters of the model, $h$, $d$, $k_1$ and $k_2$, as
\begin{equation}\label{res}
h=d\left[\frac{20k_1-k_2 \pm 4\sqrt{25k_1^2-5k_1k_2}}{40k_1+k_2}\right]^{1/2}.
\end{equation}
This dependence is illustrated in Fig.~\ref{regions}, where the currents ratio $I/i$, shown on the horizontal axis, is equivalent to $k_1/k_2$. We note, that the resonant condition (\ref{res}) implicitly accounts for the dependence on the prominence mass via the equilibrium condition (\ref{equilibrium}).

\begin{figure}
	\begin{center}
		\includegraphics[width=6cm]{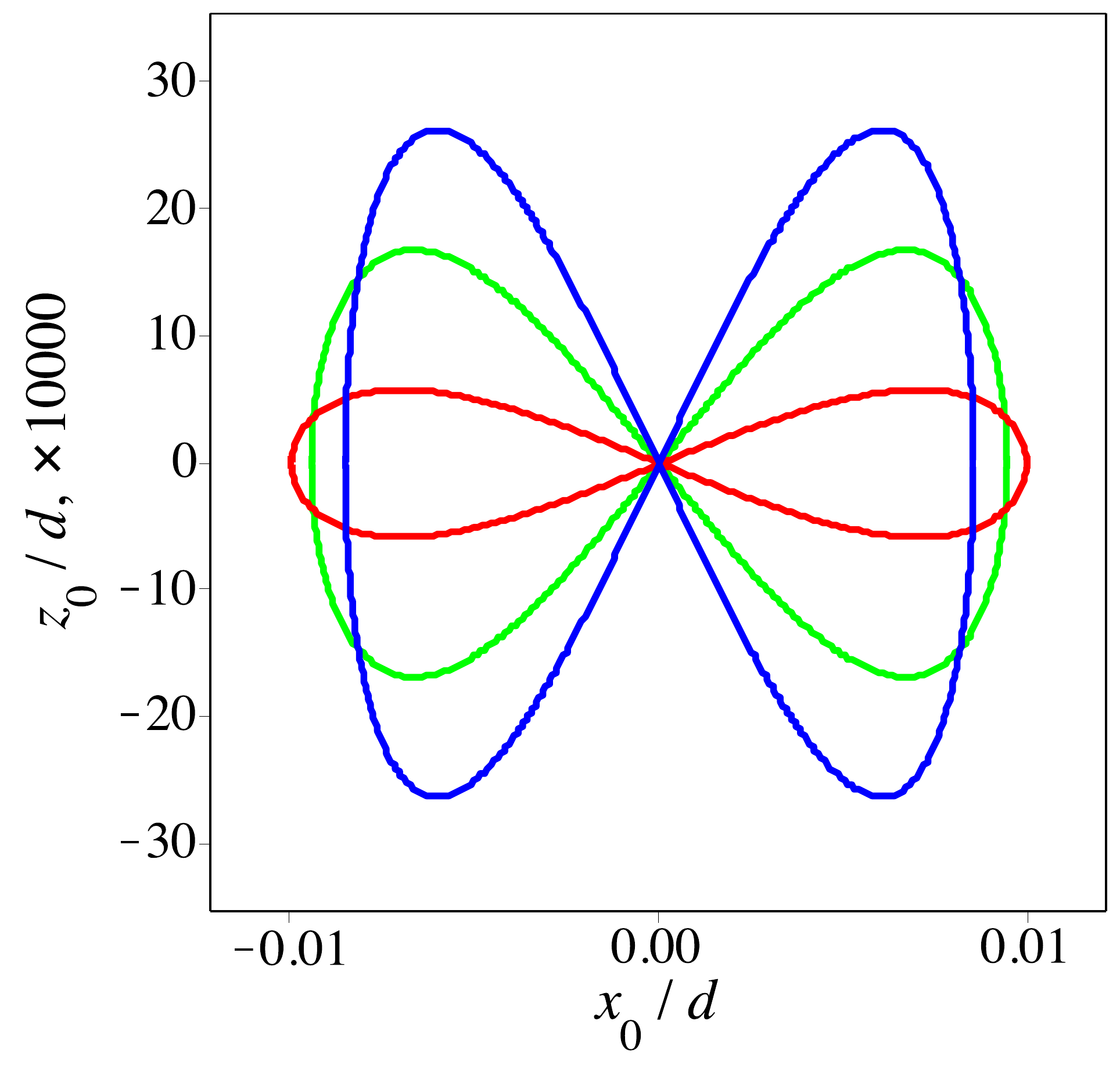}	
	\end{center}
	\caption{Displacements of the prominence in the saturated resonant nonlinear regime described by Eq.~(\ref{resz0}), shown for $h/d\approx0.244$ (see Eq.~(\ref{res})), $I/i=0.5$, $A_0/d=0.01$, and $\tau=50$ (red), 150 (green), 250 (blue), measured in units of $\sqrt{\mathcal{R} d^2/k_2}$.} 
	\label{z0fig}
\end{figure}

As long as $2\omega_1\neq\omega_2$, the prominence dynamics governed by set (\ref{Fx2})--(\ref{Fz2}), is described by solutions (\ref{solx})--(\ref{solz}). However, in the special resonant case, when $2\omega_1=\omega_2$, solutions (\ref{solx})--(\ref{solz}) break down and are no longer applicable. To describe analytically the prominence behaviour in the resonant case with $2\omega_1=\omega_2$, we introduce an additional slow time variable $\tau=\varepsilon t$ with $\varepsilon$ being a small parameter, and allow the amplitudes $A_1$ and $A_2$ in the harmonic solutions (\ref{harmonicx2})--(\ref{harmonicz2}) to be slowly varying functions of $\tau$, $A_1=A_1(\tau)$ and $A_2=A_2(\tau)$; thus $x_{0,1}(t,\tau)$ and $z_{0,1}(t,\tau)$. In such a formulation the time derivative transforms to $d/dt\equiv \partial /\partial t+\varepsilon (\partial /\partial \tau)$, and taking the initial phases to be zero one can re-write Eqs.~(\ref{nonharmonicx})--(\ref{nonharmonicz}) as
\begin{equation}\label{res_eqx}
\frac{\partial^2\,x_1}{\partial\,t^2}+\omega_1^2x_1=\frac{A_1A_2\,\beta}{2}\left\{\cos\left[(\omega_1-\omega_2)t\right]-\cos\left[(\omega_1+\omega_2)t\right] \right\}\nonumber\\
-2\frac{d\,A_1}{d\,\tau}\omega_1\cos(\omega_1t),
\end{equation}
\begin{equation}\label{res_eqz}
\frac{\partial^2\,z_1}{\partial\,t^2}+\omega_2^2z_1=\frac{A_1^2\delta}{2}+\frac{A_2^2\sigma}{2}-\frac{A_1^2\delta}{2}\cos(2\omega_1t)-\frac{A_2^2\sigma}{2}\cos(2\omega_2t)\nonumber\\
-2\frac{d\,A_2}{d\,\tau}\omega_2\cos(\omega_2t),
\end{equation}
with an additional term appearing on the right-hand side of both equations. According to solutions (\ref{solx})--(\ref{solz}), the resonance originates from the first and third terms on the right-hand side of Eqs.~(\ref{res_eqx}) and (\ref{res_eqz}), respectively. Hence, we can remove them by demanding
\begin{equation}\label{eqA1}
\omega_2\frac{d\,A_1}{d\,\tau}-\frac{A_1A_2}{2}\beta=0,
\end{equation}
\begin{equation}\label{eqA2}
2\omega_2\frac{d\,A_2}{d\,\tau}+\frac{A_1^2\delta}{2}=0,
\end{equation}
where the resonance condition $\omega_2=2\omega_1$ has been used. The set of coupled equations (\ref{eqA1})--(\ref{eqA2}) has the following solution, derived in detail in \ref{app1}:
\begin{equation}\label{solA1}
A_1=A_0\sech[A_0(\lambda)^{1/2}\tau],
\end{equation}
\begin{equation}\label{solA2}
A_2=-A_0\left(\frac{\delta}{2\beta}\right)^{1/2}\tanh[A_0(\lambda)^{1/2}\tau],
\end{equation}
where $\lambda=-\beta\delta /8\gamma$, and $A_0=A_1(0)$ is determined by the initial condition. Then substituting solutions (\ref{solA1})--(\ref{solA2}), $A_1(\tau)$ and $A_2(\tau)$, into the full expressions for the lowest order harmonic components (\ref{harmonicx2})--(\ref{harmonicz2}), one can obtain the relation between the time variations of the vertical and horizontal coordinates, $z_0$ and $x_0$ (see \ref{app1}), describing the prominence dynamics in the resonant case:
\begin{equation}\label{resz0}
z_0^2=\frac{2\delta}{\beta}x_0^2\sinh^2[A_0(\lambda)^{1/2}\tau]\left\{1-\frac{x_0^2}{A_0^2}\cosh^2[A_0(\lambda)^{1/2}\tau]\right\},
\end{equation}
with $\lambda$ and $A_0$ introduced above in Eqs.~(\ref{solA1})--(\ref{solA2}). We note that the coefficients $\beta$, $\gamma$, $\delta$, appearing in Eq.~(\ref{resz0}), are all functions of the intrinsic parameters of the model, $h$, $d$, $k_1$, and $k_2$ (see Eqs.~(\ref{Fx2})--(\ref{Fz2})), hence, their values should be chosen according to the resonant condition (\ref{res}) when operating with solution (\ref{resz0}). Prominence resonant space trajectories described by (\ref{resz0}) are illustrated in Fig.~\ref{z0fig} being Lissajous-like curves of a symmetric hourglass shape. In particular, Fig.~\ref{z0fig} clearly shows the nonlinear mode coupling effect, i.e. the increase in the vertical amplitude of the prominence oscillation with time leads to a decrease in its horizontal amplitude, thus manifesting the conservation of energy in the system.

\begin{figure*}
	\begin{center}
		\includegraphics[width=3.9cm]{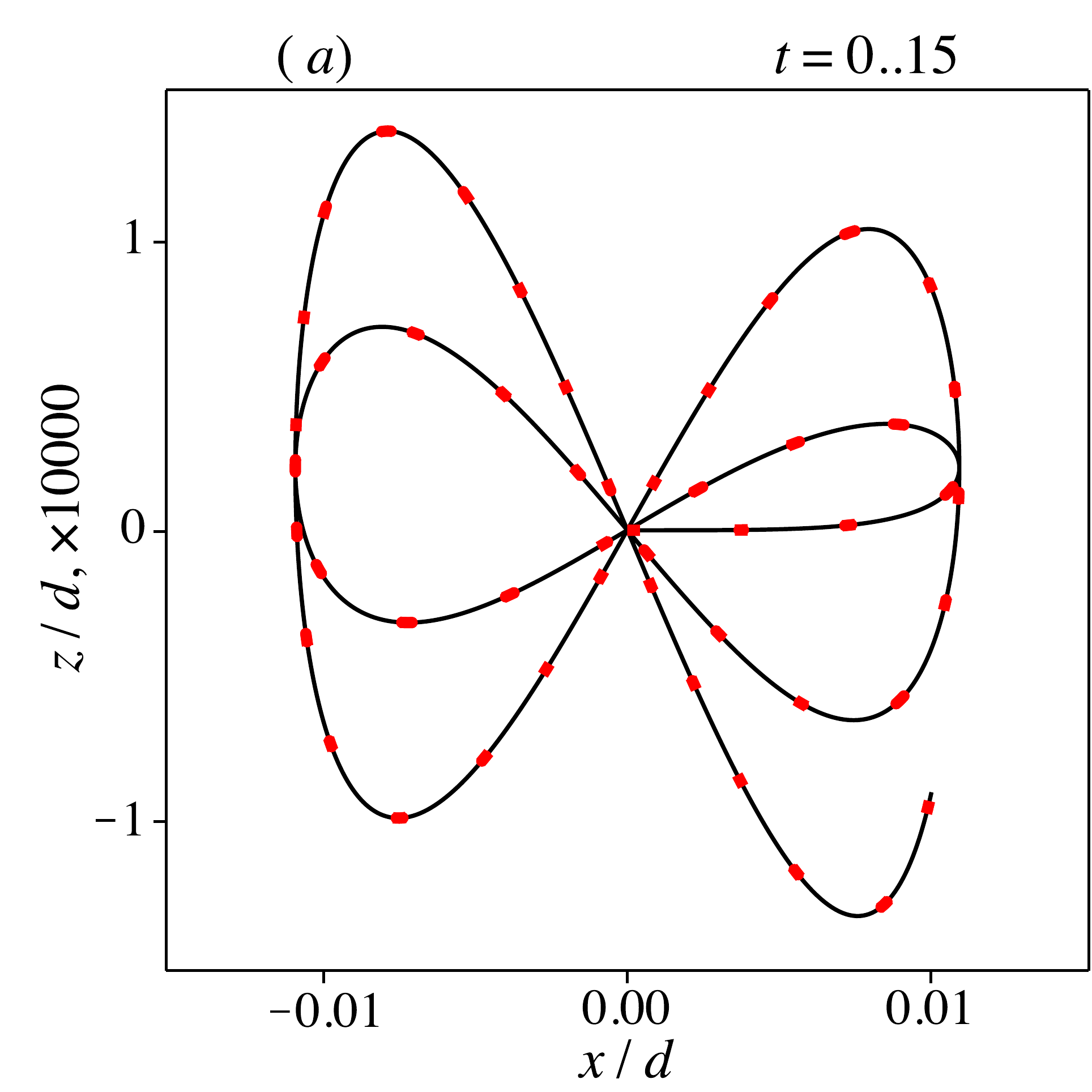}
		\includegraphics[width=3.9cm]{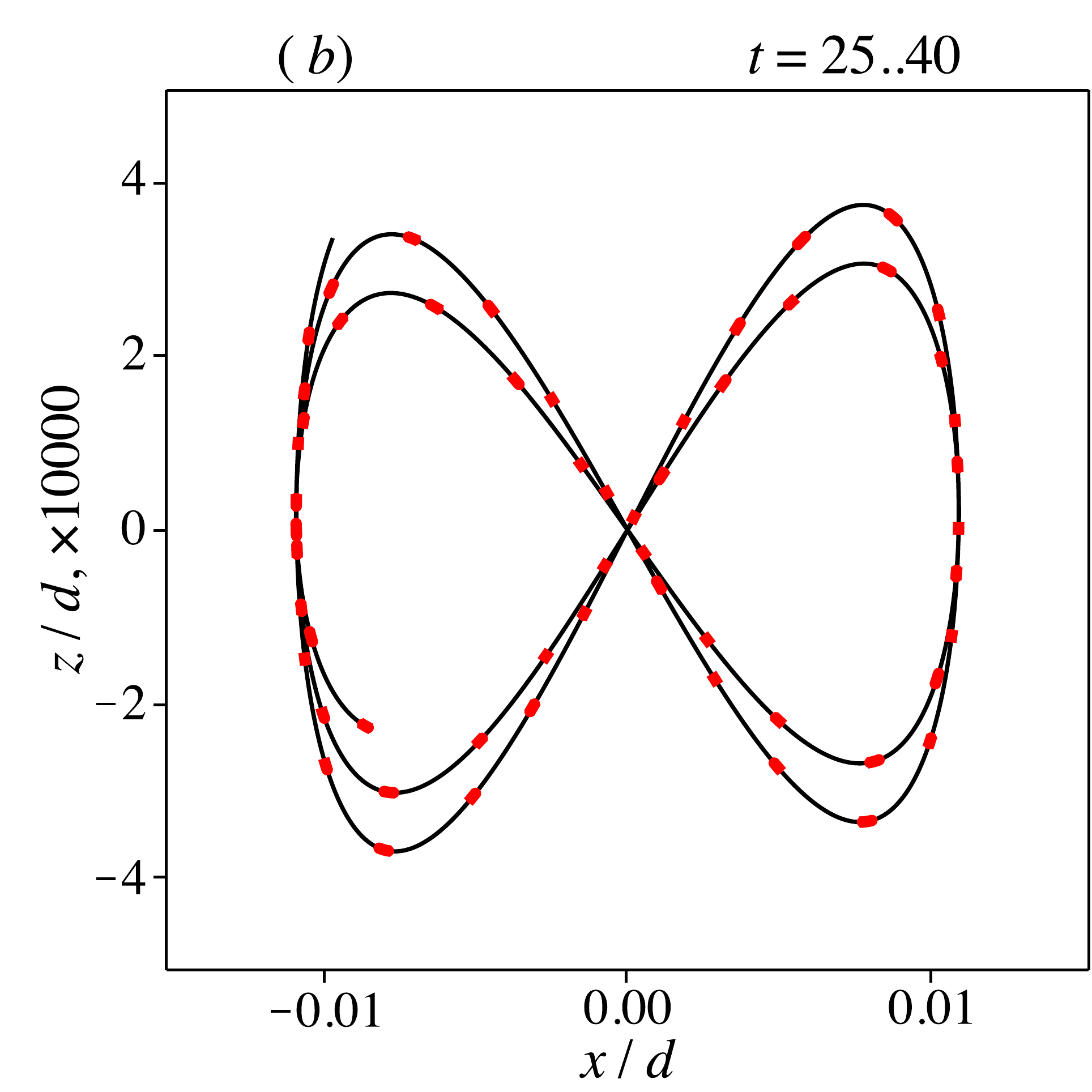}
		\includegraphics[width=3.9cm]{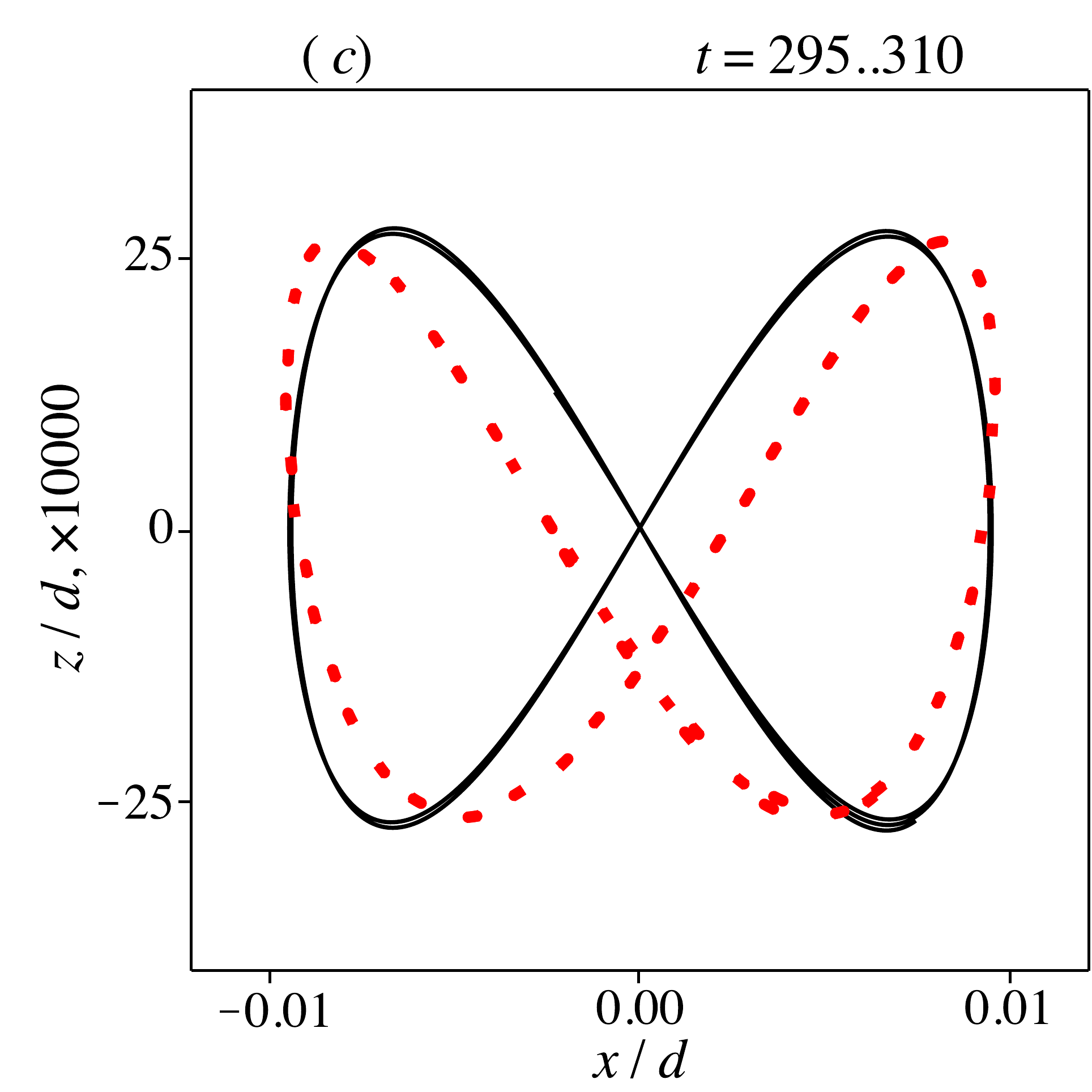}
		
		\includegraphics[width=3.9cm]{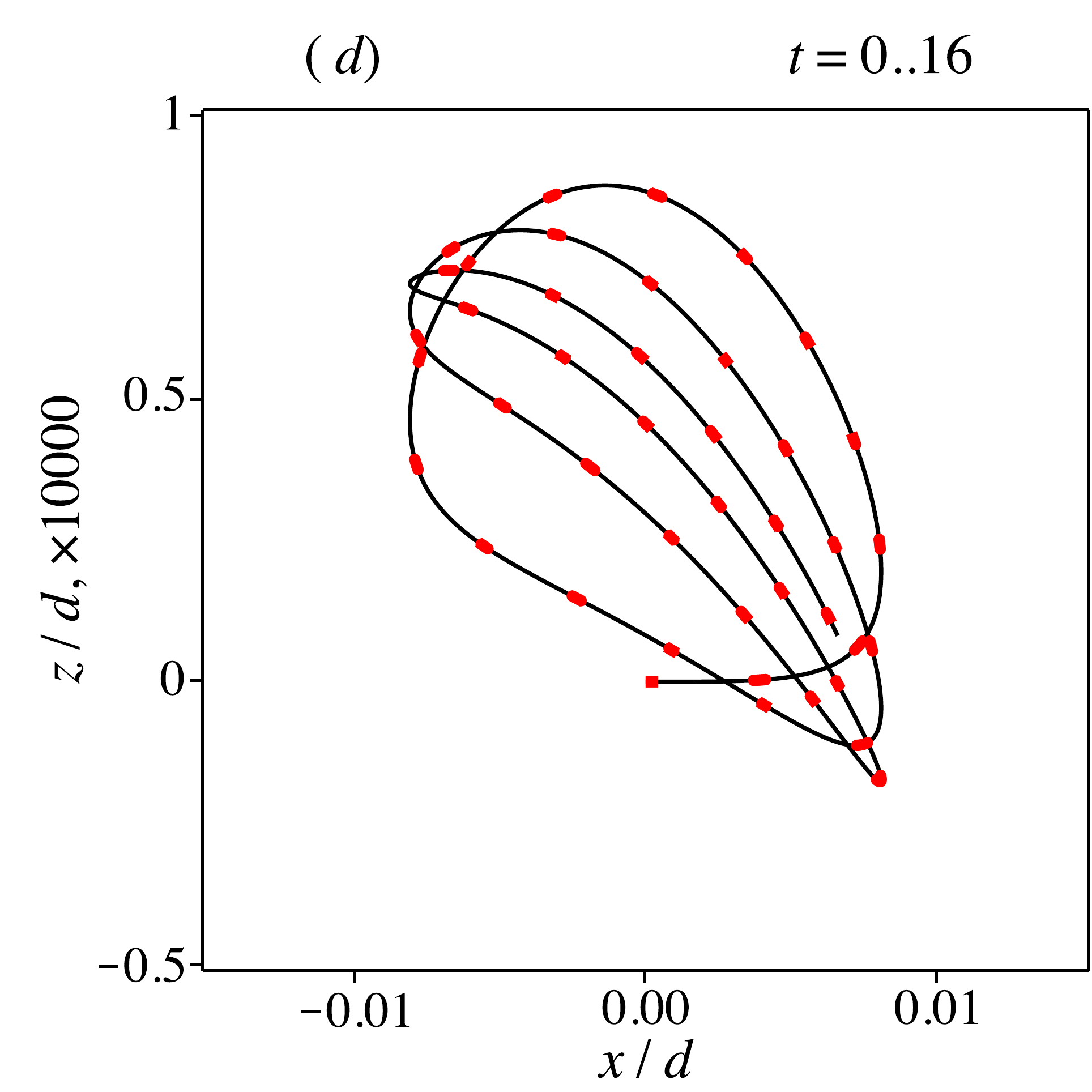}
		\includegraphics[width=3.9cm]{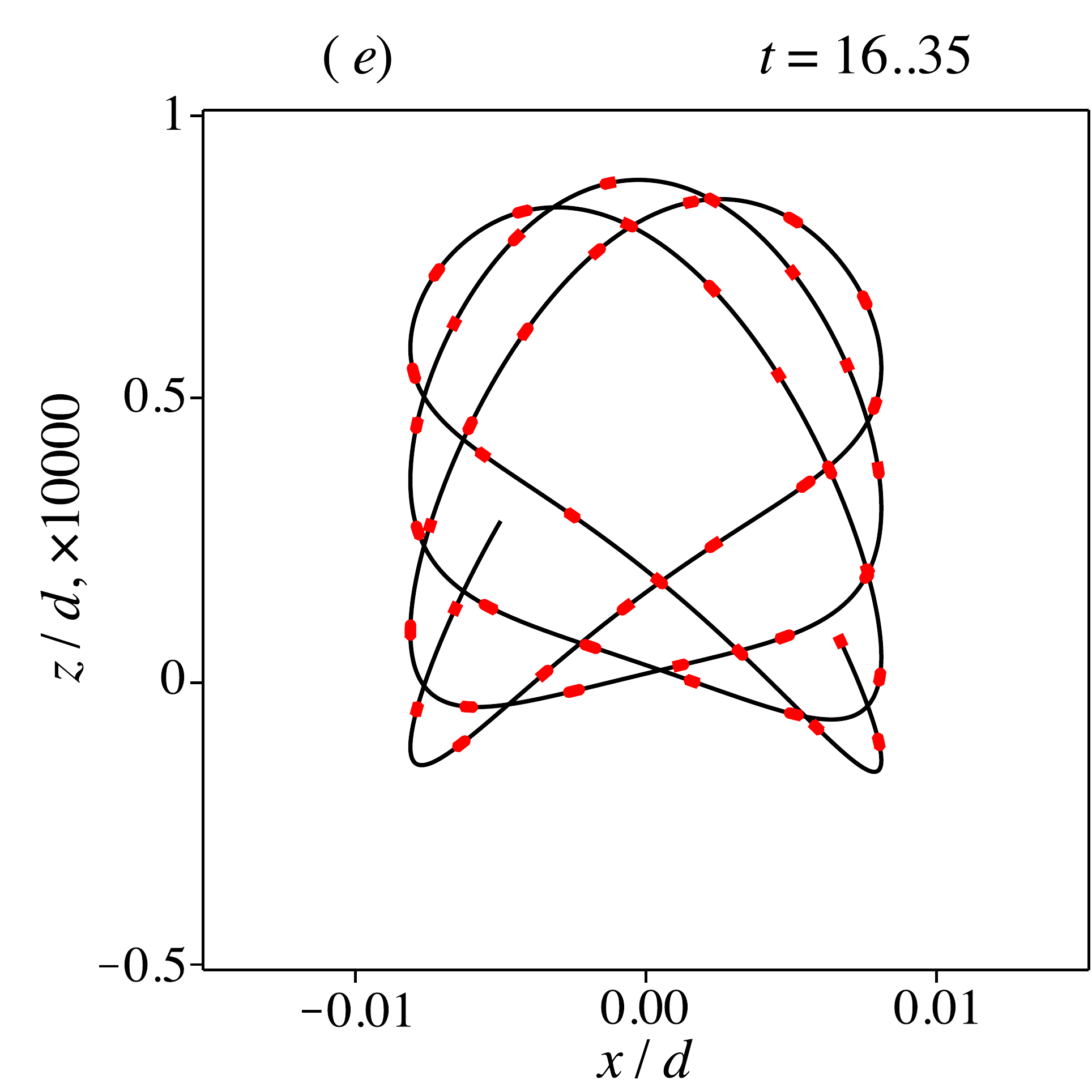}
		\includegraphics[width=3.9cm]{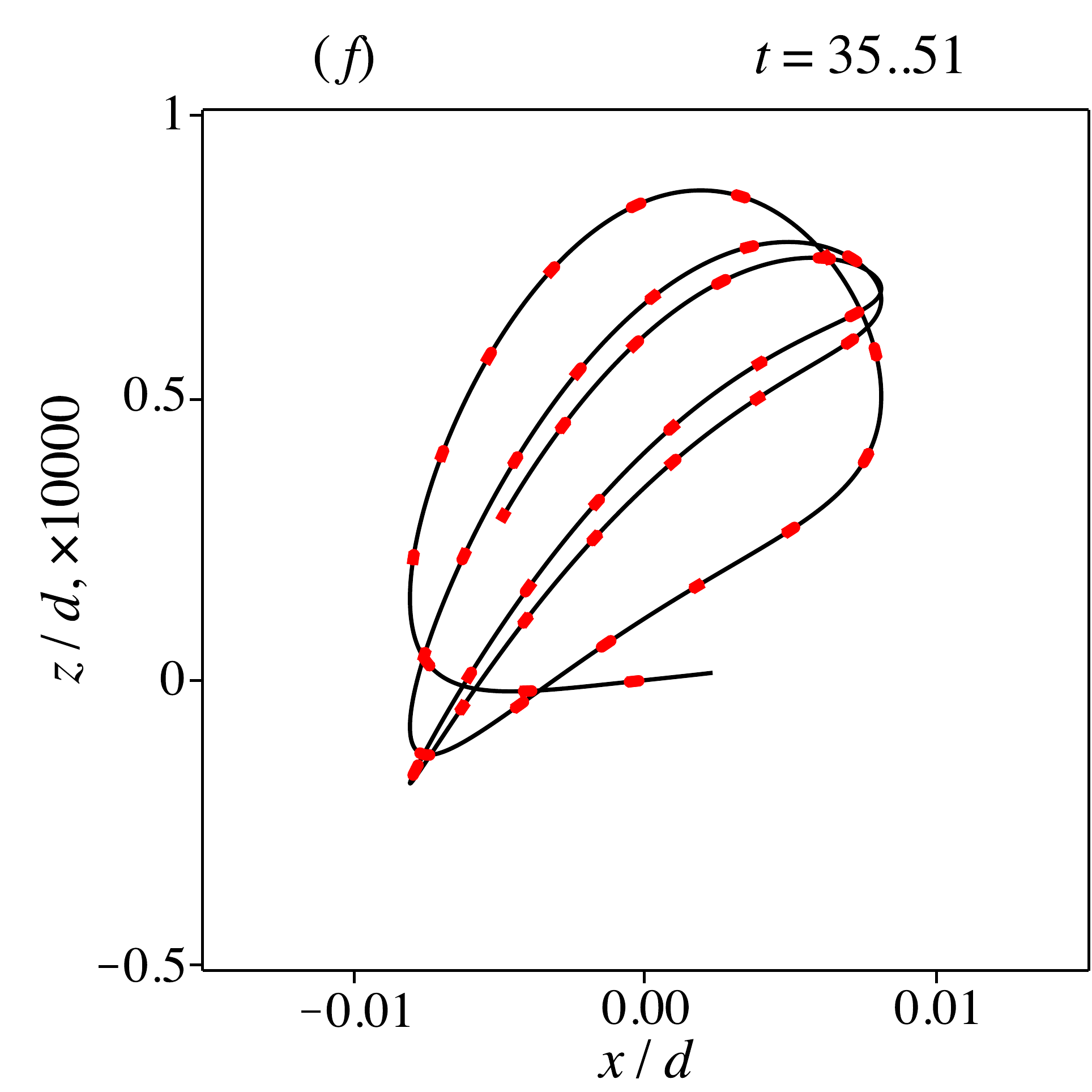}
	\end{center}
	\caption{Displacements of a current-carrying prominence in a magnetic dip during three different time intervals (shown above each panel), determined numerically as solutions of Eqs.~(\ref{Fx2})--(\ref{Fz2}) {(black solid lines) and Eqs.~(\ref{Fx})--(\ref{Fz}) (red dots),} with the initial conditions $x(0)=0$, $z(0)=0$, $\dot{x}=0.01$ (written in units of $\sqrt{k_2/\mathcal{R}}$), and $\dot{z}=0$. Panels (a)--(c) show a resonant case (see Eqs.~(\ref{res}) and (\ref{resz0}) and Figs.~\ref{regions} and \ref{z0fig}) with $h/d\approx 0.244$ and $I/i=0.5$. Panels (d)--(f) show a non-resonant case with $h/d=0.3$ and $I/i=1$. Time $t$ is measured in the units of $\sqrt{\mathcal{R} d^2/k_2}$.} 
	\label{coupfig}
\end{figure*}

\begin{figure}
	\begin{center}
		\includegraphics[width=6cm]{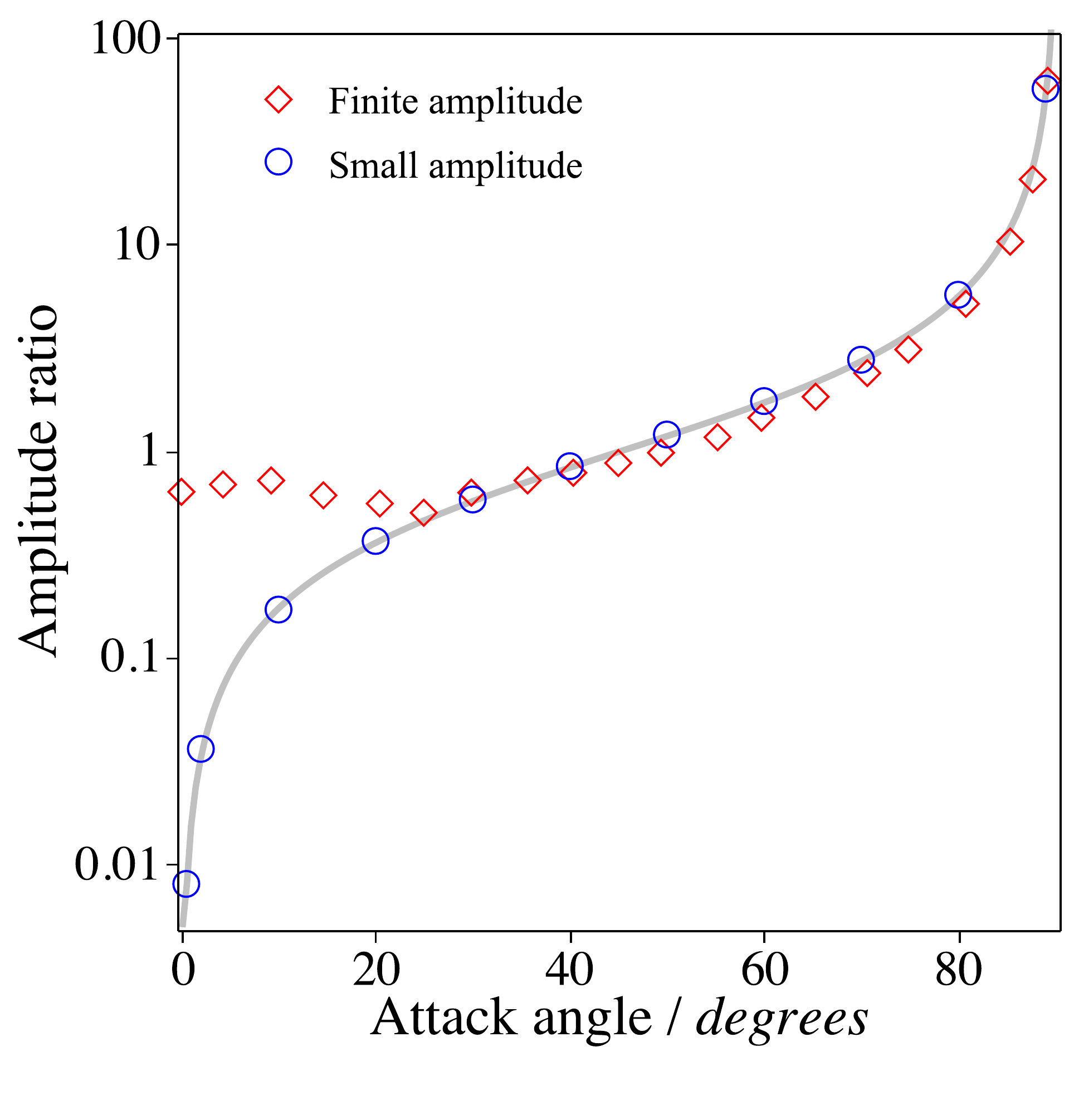}
	\end{center}
	\caption{Dependence of the maximum vertical and horizontal displacement ratio upon the angle between the direction of the initial perturbation and the horizontal axis, shown for small (blue circles) and finite (red diamonds) amplitude transverse oscillations of the prominence. The grey solid line {shows the tangent} of the attack angle. Vertical axis is shown in the logarithmic scale. The example is shown for $h/d=0.5$ and $I/i$=0.5, corresponding to a non-resonant case.} 
	\label{coup_effic_fig}
\end{figure}

The spatial polarisation of nonlinear transverse oscillations of a prominence in both resonant and non-resonant cases is shown in Fig.~\ref{coupfig}. It demonstrates the evolutionary solutions of set (\ref{Fx2})--(\ref{Fz2}), obtained with the initial conditions $x(0)=0$, $z(0)=0$, $\dot{x}=0.01$ (written in units of $\sqrt{k_2/\mathcal{R}}$), and $\dot{z}=0$, at three different time intervals of the prominence evolution. Such a set of the initial conditions implies that at the initial instant of time the prominence is located at the equilibrium position and is perturbed by a non-zero value of the horizontal speed. A possible driver is, for example, a horizontally propagating coronal wave \citep[e.g.][]{Hershaw2011, 2014ApJ...786..151S}. As the initial perturbation occurs, the prominence moves almost strictly along the $x$-axis (see panels (a) and (d) in Fig.~\ref{coupfig}) until its amplitude becomes sufficiently large (about a half of the maximum horizontal amplitude), and the vertical displacement of the prominence is generated by the nonlinear coupling mechanism described above. The latter clearly illustrates the uncoupled nature of the small-amplitude prominence oscillations considered in KNN16, and, in contrast, the highly pronounced nonlinear coupling between larger amplitude vertical and horizontal displacements. {Numerical tests {performed with the use of Eqs.~(\ref{Fx})--(\ref{Fz}) and (\ref{Fx2})--(\ref{Fz2}) solved by the 4th order Runge--Kutta scheme with the \textit{dsolve} routine in Maple 2016,} showed that the coupling works more efficiently for larger amplitude oscillations and for smaller angles between the direction of the initial perturbation and the horizontal axis (an attack angle). In the limiting case when the initial perturbation is directed strictly along the vertical axis, the set of equations (\ref{Fx2})--(\ref{Fz2}) is uncoupled for arbitrarily large oscillation amplitudes. This is illustrated by Fig.~\ref{coup_effic_fig}, where the numerical dependence of the maximum vertical and horizontal displacement ratio upon the direction of the initial perturbation is shown for small and large amplitude cases. In the case of small amplitudes, the nonlinear coupling between the vertical and horizontal modes is suppressed, and the dependence of the amplitude ratio upon the attack angle is naturally governed by a {tangent} function. In contrast, for larger amplitude displacements this dependence clearly deviates from the {tangent dependence upon the attack angle} at smaller angles of the initial perturbation, which is caused by a strong nonlinear coupling.}

In the resonant case, when the frequency of the vertical mode is twice the horizontal mode frequency, $2\omega_1=\omega_2$ (top panels of Fig.~\ref{coupfig}), horizontal displacement of the prominence achieves a nearly maximum amplitude during the first cycle of the prominence evolution (see panel (a), Fig.~\ref{coupfig}), while its vertical amplitude grows gradually, accompanied with the increasing ordering of the prominence trajectories in space. This evolution continues until the resonant limit cycle of a symmetric shape, described by Eq.~(\ref{resz0}) and shown in Fig.~\ref{z0fig}, is reached (panel (c), Fig.~\ref{coupfig}), when all trajectories are highly concentrated in space.
In contrast, in the non-resonant case (bottom panels in Fig.~\ref{coupfig}), the prominence trajectories do not experience such a localisation in space, and consequently the vertical displacement remains relatively small in amplitude in comparison with the resonant case during the whole prominence evolution. Non-resonant dynamics of the prominence can be represented by families of space trajectories, shown in panels (d) and (f), switching one to another through a transition state illustrated in panel (e). {Figure~\ref{coupfig} also shows the numerical solutions of the fully nonlinear set of equations (\ref{Fx})--(\ref{Fz}), obtained for the same values of the physical parameters of the model and initial conditions as those of set (\ref{Fx2})--(\ref{Fz2}). Both solutions are seen to be well consistent with each other justifying the analytical treatment of a non-resonant evolution of the prominence, developed in this section, except the saturated regime of the resonant case shown in panel (c). This apparent discrepancy indicates the presence of resonances also in other higher-order terms which are not accounted for by Eqs.~(\ref{Fx2})--(\ref{Fz2}). Despite these differences, the saturated resonant trajectories shown in panel (c) are seen to possess similar topologies and amplitudes, which justifies the resonant analytical solution (\ref{resz0}) too.}

\section{Fully nonlinear case}
\label{large}
\subsection{Potential energy analysis}
In this section we consider prominence oscillations, analysing Eqs.~(\ref{Fx}) and (\ref{Fz}) with the exact expressions of the forces $F_x$ and $F_z$, without usage of their Taylor expansions.
First we note that the forces $F_x$ and $F_z$ acting on the prominence in the horizontal and vertical directions, respectively, can be re-written as
\begin{equation}\label{FxD}
F_x=\frac{k_1}{2}\frac{\partial}{\partial\,x}\ln D,
\end{equation}
\begin{equation}\label{FzD}
F_z=\frac{\partial}{\partial\,z}\left[\frac{k_1}{2}\ln D+k_2\ln (2h+z)\right]-\mathcal{R} g,
\end{equation}
where
$$D\equiv (d^2-x^2)^2+2(h+z)^2(d^2+x^2)+(h+z)^4.\nonumber$$

Equations (\ref{Fx}) and (\ref{Fz}) are thus of a Hamiltonian form with $U(x,z)$ being the prominence potential energy, and $dx/dt$ and $dz/dt$ being the effective momenta. Using the relations $F_x=-\partial\,U/\partial\,x$ and $F_z=-\partial\,U/\partial\,z$, one can express the prominence effective potential energy $U(x,z)$ as
\begin{equation}\label{U}
U(x,z)=-\frac{k_1}{2}\ln D-k_2\ln (2h+z)+\mathcal{R} gz+C,
\end{equation}
where $C$ is an arbitrary constant.
\begin{figure*}
	\begin{center}
		\includegraphics[width=12.2cm]{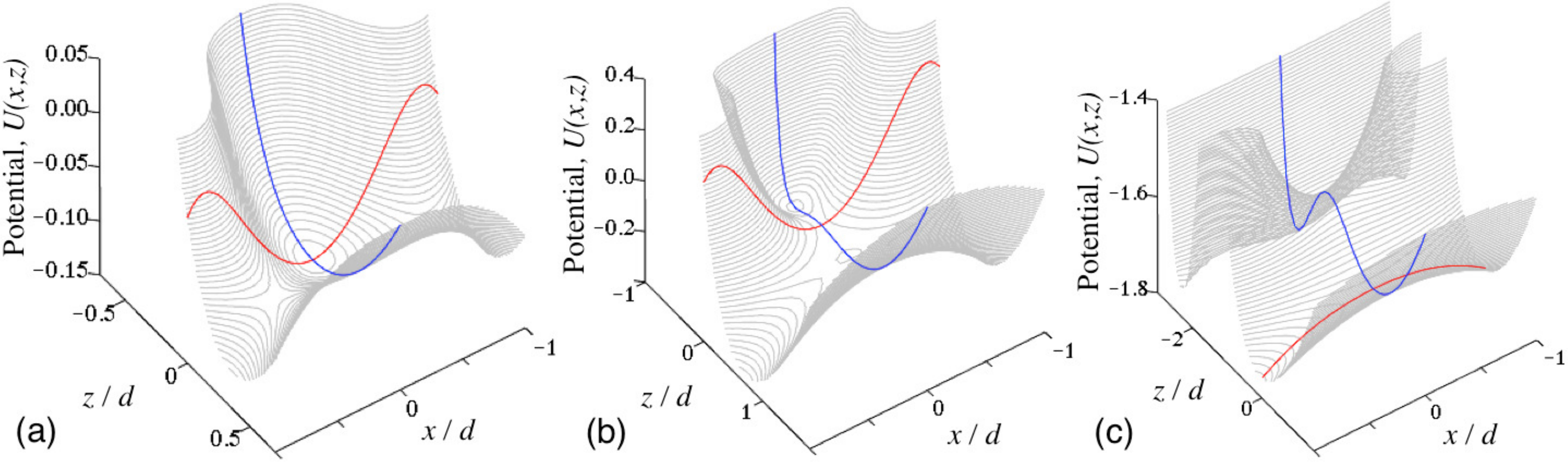}	
	\end{center}
	\caption{Potential energy $U(x,z)$ of the prominence, given by Eq.~(\ref{U}) and normalised to $k_2=\mu_0 i^2/2 \pi$. Panel (a): $h/d=0.5$, $I/i=0.5$ (region I, Fig.~\ref{regions}). Panel (b): $h/d=0.45$, $I/i=1.5$ (region II, Fig.~\ref{regions}). Panel (c): $h/d=1.5$, $I/i=0.5$ (region III, Fig.~\ref{regions}). Red and blue curves show $U(x,z=0)$ and $U(x=0,z)$ functions, respectively.}
	\label{potentials}
\end{figure*}
Behaviour of potential (\ref{U}) is shown in Fig.~\ref{potentials} for three different combinations of the intrinsic parameters of the model (i.e. $h$, $d$, $k_1$, and $k_2$), corresponding to three regions on the parametric diagram shown in Fig.~\ref{regions}.
More specifically, all panels in Fig.~\ref{potentials} sustain small amplitude decoupled prominence oscillations: in vertical and horizontal directions simultaneously (panel (a)), in the horizontal direction only (panel (b)), and in the vertical direction only (panel (c)), which is consistent with the linear theory developed in KNN16.
In the nonlinear case, panel (a) shows the potential surface $U(x,z)$ with a local dip of a finite height, corresponding to a locally stable (or metastable) equilibrium of the prominence. Such a metastable prominence state allows for the essentially coupled nonlinear oscillations with a critical amplitude, above which the prominence becomes horizontally unstable.
In turn, nonlinear large amplitude oscillations in the cases shown in panels (b) and (c) may quickly become unstable in the horizontal direction due to the nonlinear coupling mechanism described in the previous section.

\subsection{Maximum horizontal and vertical amplitudes}
We now investigate the dependence of the maximum oscillation amplitudes in a metastable prominence state upon the intrinsic physical parameters of the model ($h$, $d$, $k_1$, and $k_2$) addressing the potential energy example shown in Fig.~\ref{potentials}, panel (a) with $h<d$ and $k_1<k_2$. For that we analyse the positions $x_m$ and $z_c$ of the local extrema of the function $U(x,z)$ (\ref{U}) by solving the following set of coupled equations
\begin{equation}\label{Fxc}
F_x(x_m,z_c)=0,
\end{equation} 
\begin{equation}\label{Fzc}
F_z(x_m,z_c)=0,
\end{equation} 
where $F_x$ and $F_z$ are the forces given in Eqs.~(\ref{Fx}) and (\ref{Fz}), respectively. In addition to the trivial solution of set (\ref{Fxc})--(\ref{Fzc}) with $x_m=0$ and $z_c=0$, corresponding to the initial equilibrium of the prominence, another real solution in the region of parameters $h<d$ and $k_1<k_2$ is possible:
\begin{eqnarray}\label{zc}
z_c=\frac{(1/2)\,h}{k_2(h^2+d^2)+4h^2k_1}\bigg\{2k_1(d^2-5h^2)-k_2(h^2+d^2) \nonumber\\
+\sqrt{[k_2(d^2+h^2)+2k_1(d^2+3h^2)]^2+8k_1k_2(d^4-h^4)}\bigg\},
\end{eqnarray} 
\begin{eqnarray}\label{xm}
|x_m|=\Bigg\{d^2-\frac{(1/4)\,h^2}{(k_2(h^2+d^2)+4h^2k_1)^2}\bigg[k_2(h^2+d^2)+2k_1(d^2-h^2)\nonumber \\
+\sqrt{[k_2(d^2+h^2)+2k_1(d^2+3h^2)]^2+8k_1k_2(d^4-h^4)}\bigg]^2\Bigg\}^{1/2}.
\end{eqnarray}

\begin{figure}
	\begin{center}	
		\includegraphics[width=9cm]{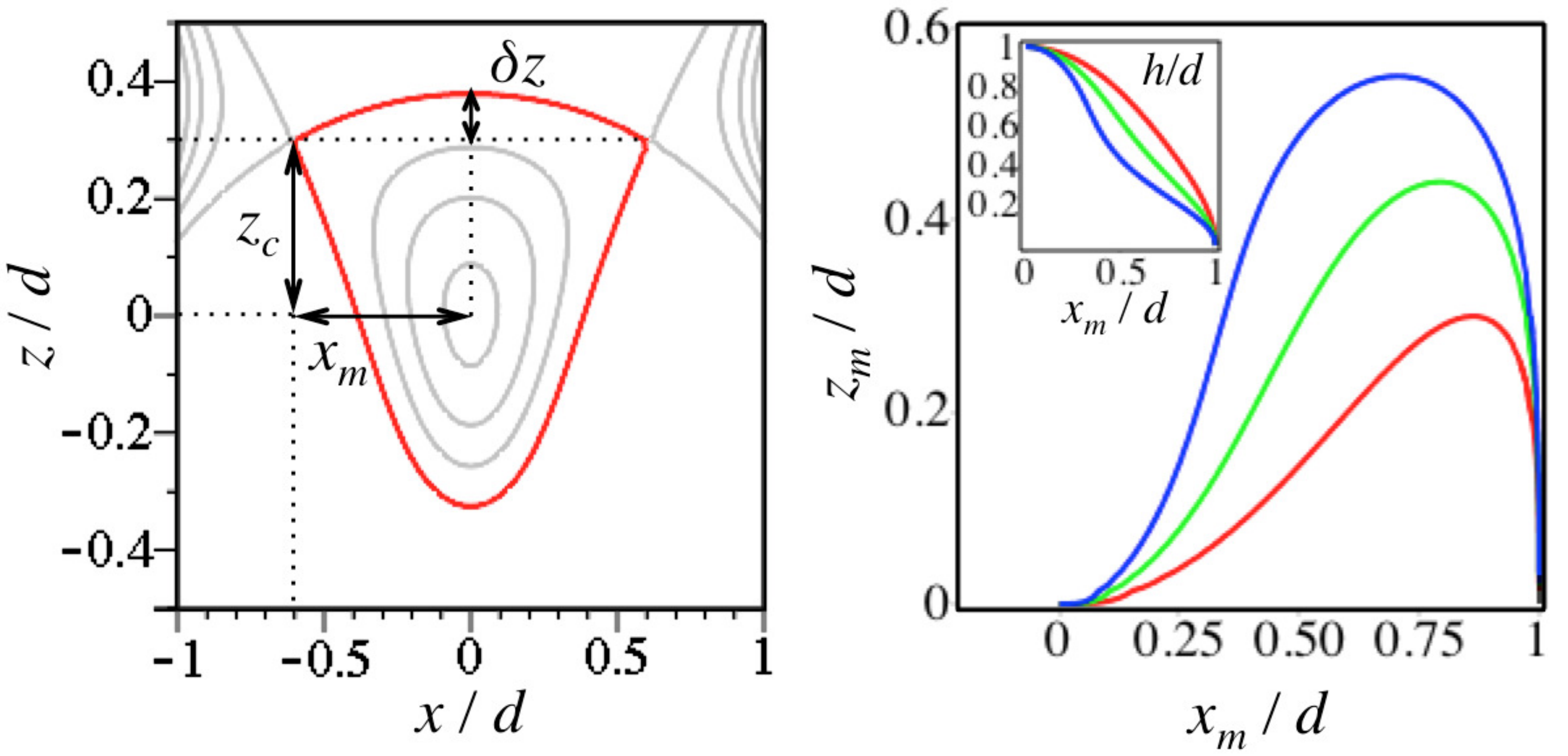}	
	\end{center}
	\caption{\textit{Left}: space contours showing the equipotential levels of the prominence potential energy shown in Fig.~\ref{potentials}, panel (a), up to the critical value shown in red and determined by $U_c=U(x_m,z_c)$, with $U(x,z)$, $x_m$, and $z_c$ given in Eqs.~(\ref{U}), (\ref{xm}), and (\ref{zc}), respectively.
		\textit{Right}: parametric plot showing the dependence of the maximum vertical amplitude, $z_m$ of the prominence oscillation upon its maximum horizontal amplitude, $x_m$ through the parameter $h/d$ varying from 0 to 1 and shown in the left top corner (see {Eqs.~(\ref{xm}) and (\ref{zm})}). The dependences are shown for $k_1/k_2=0.9$ (blue), $k_1/k_2=0.5$ (green), and $k_1/k_2=0.2$ (red).}
	\label{maxamp}
\end{figure}

The critical value $U_c$ of the prominence potential energy, corresponding to these $x_m$ and $z_c$, can be found as $U_c=U(x_m,z_c)$ with the function $U(x,z)$ given in Eq.~(\ref{U}). This critical value $U_c$ describes the highest prominence potential energy, above which the prominence has enough energy to escape the potential dip, becoming unstable in the horizontal direction. Figure \ref{maxamp}, left panel illustrates the equipotential levels corresponding to the closed contours in the $(x,z)$--plane, including the critical value $U_c$ with the critical space contour shown in red. According to the left panel of Fig.~\ref{maxamp}, the horizontal coordinate of the potential energy local extrema always shows the maximum possible horizontal amplitude $x_m$ allowing for the stable large amplitude prominence oscillations with energies being below the critical value of $U_c$. However, because of the vertical asymmetry of the prominence potential energy (directly connected to the vertical asymmetry of the whole model, see Fig.~1 in KNN16), the corresponding critical value of the vertical coordinate, $z_c$, in general may represent not the highest possible vertical oscillation amplitude. The latter, in turn, can be represented as (see Fig.~\ref{maxamp}):
\begin{equation}\label{zm}
z_m=z_c+\delta z,
\end{equation}
and implicitly determined by the condition $U(x=0,z_m)=U_c$, with $U_c=U(x_m,z_c)$ and  $z_c$ and $x_m$ given in Eqs.~(\ref{zc})--(\ref{xm}).

In the limit $d\gg h$, when the external photospheric currents are located at sufficiently large but finite distances from the prominence position, and hence the magnetic dip is significantly shallowed, the maximum horizontal oscillation amplitude $x_m$ can be estimated by the lowest order expansion with respect to a small parameter $h/d$, as $x_m\approx d$. This coincides with the case considered in \citet{KR1974}, where $d$ tends to infinity and the magnetic dip is completely degenerated, therefore horizontal prominence oscillations are essentially impossible and vertical oscillations may have large amplitudes limited by the height of the prominence above the surface of the Sun only. In contrast, our model supports oscillations in both directions simultaneously, and the appearance of a maximum vertical amplitude, $z_m$ given in (\ref{zm}), is attributed to the nonlinear coupling between the horizontal and vertical modes.

\begin{figure}
	\begin{center}
		\includegraphics[width=9cm]{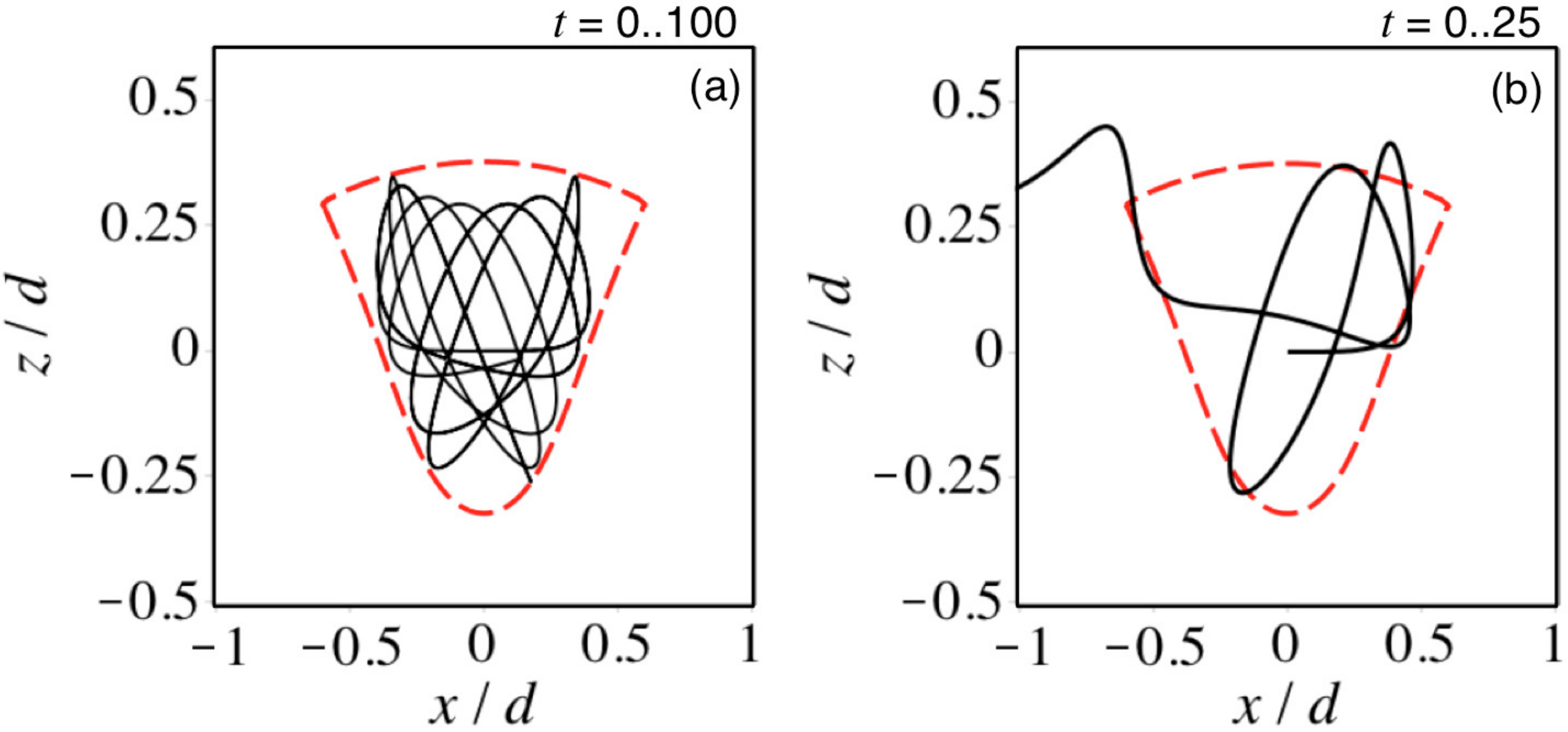}
	\end{center}
	\caption{Displacements of the prominence in the large amplitude oscillatory regime, governed by Eqs.~(\ref{Fx})--(\ref{Fz}) with $h/d=0.5$ and $k_1/k_2=0.5$, obtained with the initial conditions: $x(0)=0$, $z(0)=0$, $\dot{z}=0$, and $\dot{x}=0.26$ (in units of $\sqrt{k_2/\mathcal{R}}$, panel (a)) and $\dot{x}=0.3$ (same units, panel (b)). Time $t$ is measured in units of $\sqrt{\mathcal{R} d^2/k_2}$.}
	\label{num_sol}
\end{figure}

Dependence of the maximum horizontal and vertical amplitudes, $x_m$ and $z_m$ given by Eqs.~(\ref{xm})--(\ref{zm}) upon the intrinsic parameters of the model, including the limiting case $d\gg h$, is illustrated in Fig.~\ref{maxamp}, right panel.
In contrast to the horizontal maximum amplitude $x_m$ (\ref{xm}), which is a monotonically decreasing function of the prominence height above the photosphere, $h$, the vertical maximum amplitude $z_m$ (\ref{zm}) has a maximum at a certain value of $h$.
For example, for a nearly equal photospheric and prominence currents, $I=0.9\,i$ (blue lines), the highest value of the maximum vertical amplitude appears for $h\approx 0.28\,d$ and is about $0.55\,d$, which forces the horizontal critical amplitude to be about $0.7\,d$.

\begin{figure}
	\begin{center}
		\includegraphics[width=5.8cm]{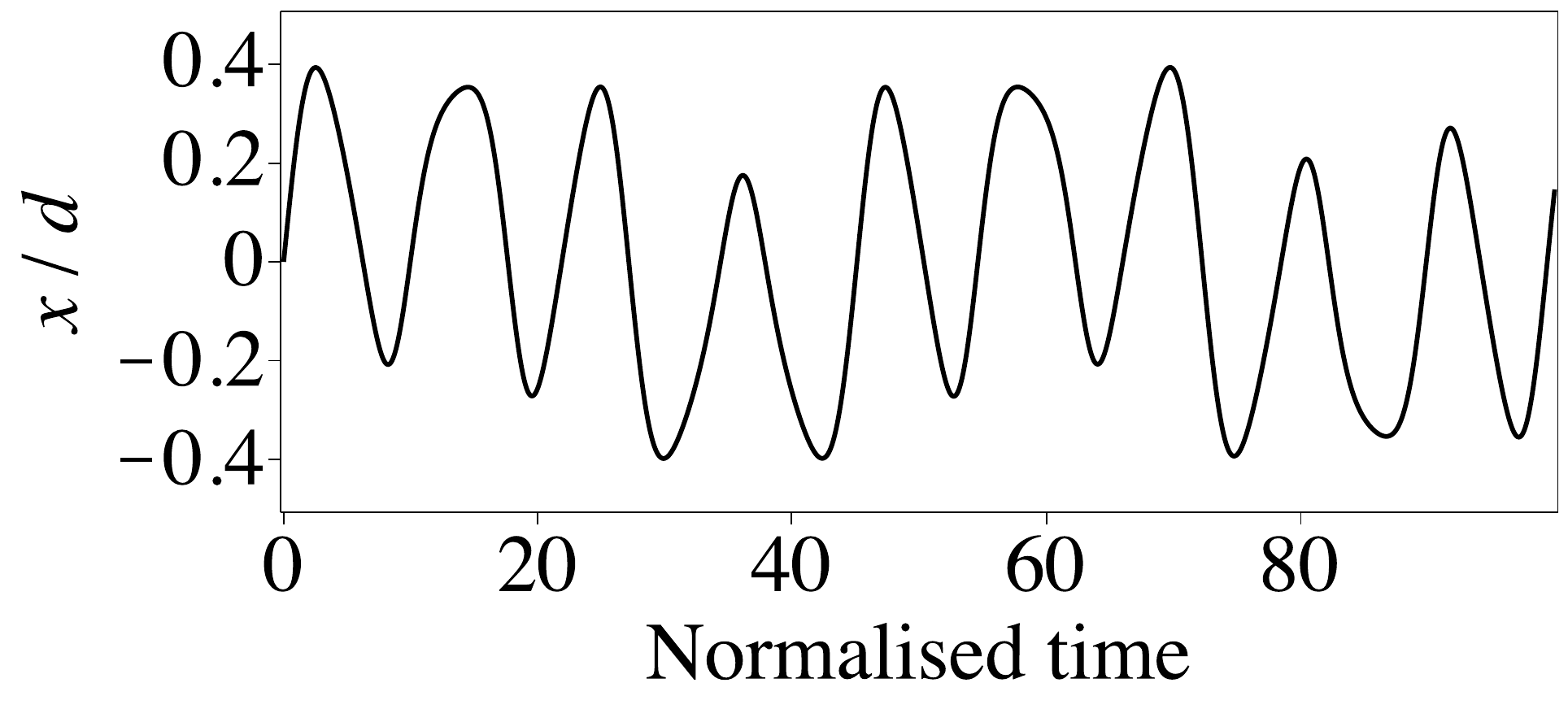}
		\includegraphics[width=5.8cm]{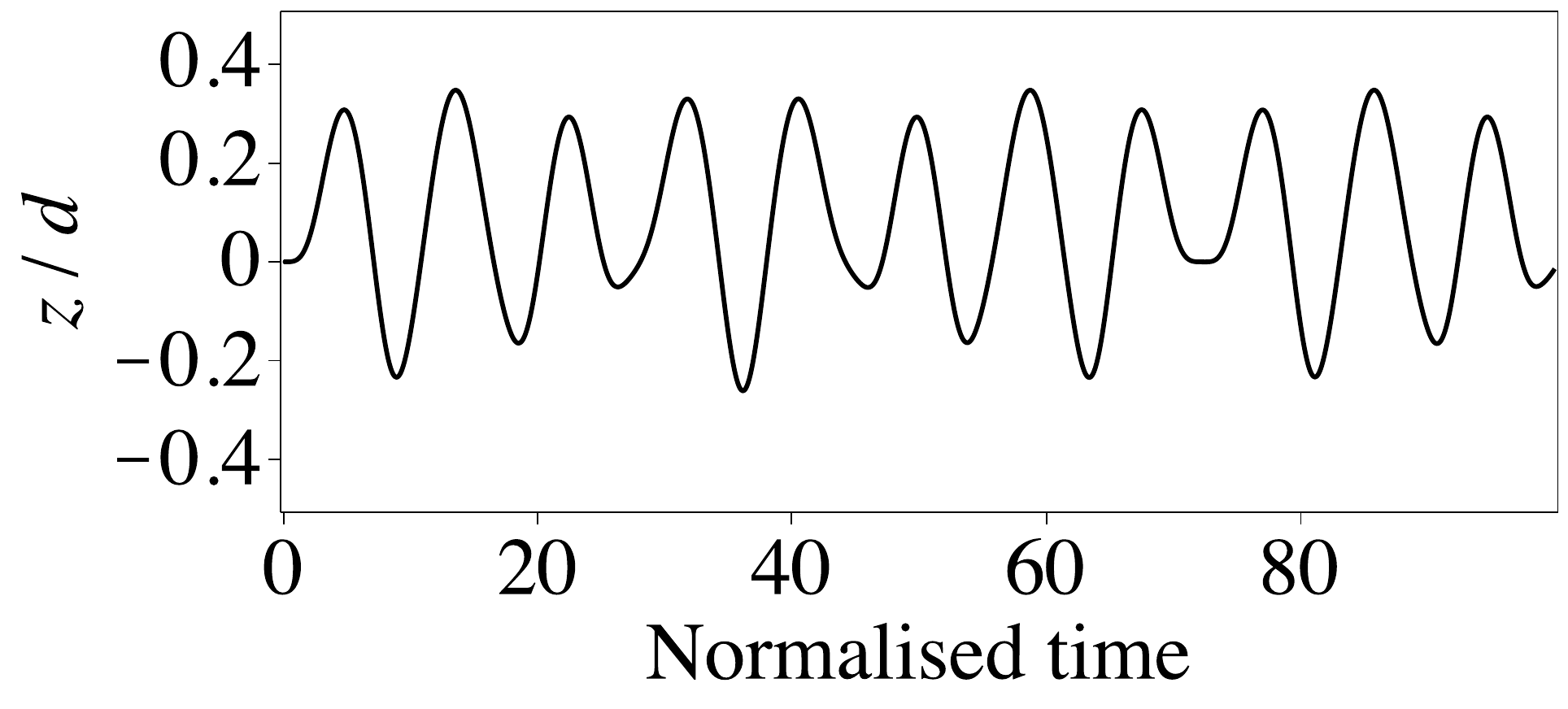}
	\end{center}
	\caption{{Temporal quasi-periodic variations of the horizontal (left) and vertical (right) displacements of the prominence in the large amplitude oscillatory regime shown in the left-hand panel of Fig.~\ref{num_sol}.
			Time is normalised to $\sqrt{\mathcal{R} d^2/k_2}$.}
	}
	\label{num_sol_osc}
\end{figure}

Figure \ref{num_sol} shows the spatial structure of large amplitude transverse oscillations of the prominence, determined by the solution of the full set (\ref{Fx})--(\ref{Fz}). Panel (a) illustrates the case when the prominence energy is slightly lower than the critical value of $U_c$, all amplitudes are always restricted by the maximum displacement (shown in red), corresponding to $U_c$, and hence the oscillations are always stable. Another case is shown in panel (b), when the prominence energy is slightly greater than $U_c$. In this regime oscillation amplitudes may exceed the critical values, which forces the prominence to become horizontally unstable in a few oscillation cycles. {We need to mention that in the stable regime shown in panel (a) of Fig.~\ref{num_sol}, the horizontally and vertically polarised modes are not strictly periodic, but could be considered as quasi-periodic with a relatively stable oscillation period and slowly modulated amplitude (see Fig.~\ref{num_sol_osc}).}

\subsection{Periods of nonlinear oscillations}
In this section we estimate analytically typical periods of finite amplitude nonlinear transverse oscillations in both horizontal and vertical directions. For this we use the conditions $z=0$ and $x=0$ in the equations of motion (\ref{Fx}) and (\ref{Fz}), respectively. In order to avoid the prominence instability caused by the nonlinear mode coupling, we restrict the oscillation amplitudes in both directions to be lower than or equal to $x_m$ (\ref{xm}) and $z_m$ (\ref{zm}). The latter means that the prominence oscillates strictly inside the potential dip shown in Fig.~\ref{potentials}, panel (a), and hence the oscillations are always stable.

Substituting $z=0$ in the equation of motion along the horizontal axis (\ref{Fx}), one can obtain its first integral representing the prominence's conservation energy law in the horizontal direction,
\begin{equation}\label{energyx}
\frac{\mathcal{R}}{2}\left(\frac{d\,x}{d\,t}\right)^2+U_x(x)=\mathcal{E}_x,
\end{equation}
where
$$U_x=-\frac{k_1}{2} \ln\left[\frac{(d^2-x^2)^2+2(d^2+x^2)h^2+h^4}{(d^2+h^2)^2}\right]\nonumber$$
is the prominence potential energy in the horizontal direction, which can be derived from  Eq.~(\ref{U}) in the limit $z=0$. The constant $\mathcal{E}_x$ in Eq.~(\ref{energyx}) shows the total energy of horizontal oscillations and can be obtained from the initial conditions $\dot{x}(0)=0$ and $x(0)=A_x$, as 
$$\mathcal{E}_x=-\frac{k_1}{2} \ln\left[\frac{(d^2-A_x^2)^2+2(d^2+A_x^2)h^2+h^4}{(d^2+h^2)^2}\right],\nonumber$$
with $A_x$ being the horizontal oscillation amplitude.

The period $P_x$ of the horizontal oscillations of an arbitrary amplitude as a function of the oscillation amplitude $A_x$ and intrinsic parameters of the model, can be derived from Eq.~(\ref{energyx}) as
\begin{equation}\label{nperiodx}
P_x=4\sqrt{\mathcal{R}}\int_0^{A_x}\frac{d\,x}{\sqrt{2(\mathcal{E}_x-U_x)}},
\end{equation}
where the functions $U_x$ and $\mathcal{E}_x$ are determined in Eq.~(\ref{energyx}).

Similarly to the horizontal case, the equation of motion along the vertical axis (\ref{Fz}), integrated once, reduces to the vertical conservation energy law
\begin{equation}\label{energyz}
\frac{\mathcal{R}}{2}\left(\frac{d\,z}{d\,t}\right)^2+U_z(z)=\mathcal{E}_z,
\end{equation}
where the vertical potential energy, obtained from Eq.~(\ref{U}) with $x=0$, is
$$U_z=-\frac{k_1}{2} \ln\left[\frac{d^4+2d^2(h+z)^2+(h+z)^4}{(d^2+h^2)^2}\right]-k_2\ln\left[\frac{2h+z}{2h}\right]+\mathcal{R} gz,\nonumber$$
and the total vertical oscillation energy can be determined from the initial conditions $\dot{z}(0)=0$ and $z(0)=A_z$, with $A_z$ being the vertical oscillation amplitude, as
$$\mathcal{E}_z=-\frac{k_1}{2} \ln\left[\frac{d^4+2d^2(h+A_z)^2+(h+A_z)^4}{(d^2+h^2)^2}\right]-k_2\ln\left[\frac{2h+A_z}{2h}\right]\nonumber\\
+\mathcal{R} gA_z.\nonumber$$
We note that the vertical equilibrium condition (\ref{equilibrium}) can be used in the above expressions for $U_z$ and $\mathcal{E}_z$ to re-write the gravitational term $\mathcal{R} g$ in terms of $h$, $d$, $k_1$, and $k_2$.
The subsequent integration of Eq.~(\ref{energyz}) allows us to derive the dependence of the arbitrarily large amplitude vertical oscillation period upon the parameters of the model and the vertical oscillation amplitude $A_z$ as
\begin{equation}\label{nperiodz}
P_z=4\sqrt{\mathcal{R}}\int_0^{A_z}\frac{d\,z}{\sqrt{2(\mathcal{E}_z-U_z)}},
\end{equation}
with the functions $U_z$ and $\mathcal{E}_z$ given above in Eq.~(\ref{energyz}).

\begin{figure}
	\begin{center}
		\includegraphics[width=8cm]{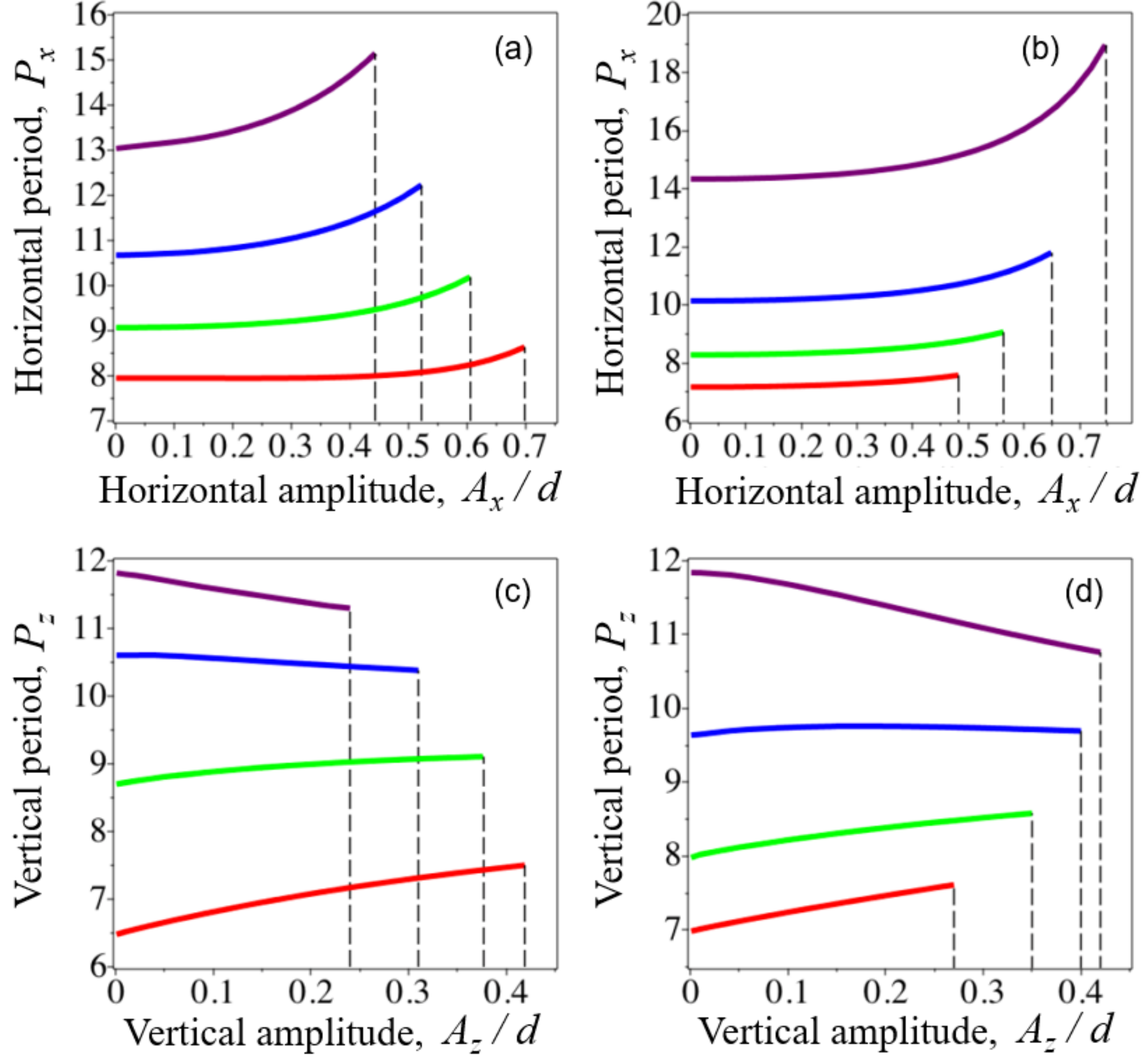}	
	\end{center}
	\caption{Dependences of horizontal and vertical periods $P_x$ (\ref{nperiodx}) and $P_z$ (\ref{nperiodz}) upon the corresponding amplitudes, $A_x$ and $A_z$, shown for different sets of the equilibrium parameters of the model. Panel (a):  $h/d=$~0.7 {(purple)}, 0.6 {(blue)}, 0.5 {(green)}, 0.4 {(red)}; $k_1/k_2=0.5$. Panel (b): $h/d=0.5$; $k_1/k_2=$~0.2 (purple), 0.4 {(blue)}, 0.6 {(green)}, 0.8 {(red)}. Panel (c): $h/d=$~0.7 {(purple)}, 0.6 {(blue)}, 0.5 {(green)}, 0.4 {(red)}; $k_1/k_2=0.5$. Panel (d): $h/d=0.5$; $k_1/k_2=$~0.75 {(purple)}, 0.6 {(blue)}, 0.4 {(green)}, 0.2 {(red)}. Vertical dashed lines indicate the maximum possible amplitudes, $x_m$ and $z_m$ , determined by Eqs.~(\ref{xm}) and (\ref{zm}), respectively, in each case. Periods $P_z$ and $P_x$ are measured in the units of $\sqrt{\mathcal{R} d^2/k_2}$.
	}
	\label{periods}
\end{figure}

Dependences of the horizontal and vertical oscillation periods, $P_x$ (\ref{nperiodx}) and $P_z$ (\ref{nperiodz}) upon the oscillation amplitudes $A_x$ and $A_z$, respectively, are illustrated in Fig.~\ref{periods} for different sets of the equilibrium parameters of the model, taken in the range $h<d$ and $k_1<k_2$. In these examples the amplitudes $A_z$ and $A_x$ were additionally restricted by the maximum values of $z_m$ (\ref{zm}) and $x_m$ (\ref{xm}), respectively, corresponding to each particular set of parameters. The latter guarantees the prominence oscillations being always stable, even in the case of strong coupling between horizontal and vertical modes. More specifically, in the limiting case of small amplitudes, the periods in all panels are nearly constant, which coincides with the linear theory results obtained in KNN16, where the oscillations were found to be isochronous, i.e. the oscillation periods are independent of the oscillation amplitude. In contrast, in the nonlinear large amplitude regime the horizontal period $P_x$ was found to be always increasing with the amplitude $A_x$ (panels (a) and (b) in Fig.~\ref{periods}), with the highest increase appearing for larger values of $h/d$ (panel (a)) and lower values of $k_1/k_2$ (or $I/i$, panel (b)). The dependence of the vertical period $P_z$ upon the vertical amplitude $A_z$ in the nonlinear case shows rather different behaviour (panels (c) and (d)). Namely, it increases with the amplitude for lower values of $h/d$ and $k_1/k_2$, and then changes its gradient to a negative one for higher values of these two parameters, through a transient state (green and blue lines in panels (c) and (d), respectively), where the periods are nearly constant for all allowed amplitudes. However, the period $P_z$ was detected to be weakly dependent upon the amplitude $A_z$. Indeed, the nonlinear relative change of the vertical period $P_z$ with the amplitude is of about several percent only for all examples shown in panels (c) and (d).

\section{Summary of results and conclusions}
We studied analytically finite amplitude transverse oscillations of massive quiescent current-carrying prominences in a magnetic field dip, representing a synthesis of the Kippenhahn--Schl\"uter \citep[][]{KS1957} and Kuperus--Raadu \citep[][]{KR1974} models. The model accounts for the effect of a non-zero value of the electric current in the prominence, and is based upon the electromagnetic interaction between the prominence current and the external photospheric currents producing a magnetic dip.
We derived and analysed the equations of motion in the horizontal and vertical directions, (\ref{Fx2})--(\ref{Fz2}) for weakly nonlinear oscillations, which account for the effects of the quadratic nonlinearity. Also, we studied the fully nonlinear exact set of the governing equations in both directions, (\ref{Fx})--(\ref{Fz}). {Dissipative effects such as resonant absorption, aerodynamic friction, viscosity and resistivity, and effects of partial ionisation are neglected in this work. It allows us to perform an analytical study of the nonlinear effects, including mode coupling, on the oscillations. Even in the case of effective dissipation, our results are important for understanding the initial stage of the oscillations.}

{The nonlinear oscillatory dynamics of the prominence is determined by the oscillation amplitude and was found to be highly sensitive to the parameters of the equilibrium: the value of the prominence current, its mass and position above the photosphere, and the properties of the magnetic dip. In contrast to the other parameters, that can be obtained from observations, the prominence current is not a directly observable quantity \citep[e.g.][]{2010SSRv..151..333M}. However, its value could be evaluated indirectly. For example, based on the analysis of vector magnetograms, vertical electric currents in a near-sunspot environment were found to be about $10^{11}$\,A \citep{1964SSRv....3..451S}, and of about $4\times 10^{10}$\,A in a flaring region \citep{2014ApJ...788L..18S}. Similar values of the electric currents should exist in the corona, in particular in the magnetic flux ropes of prominences. For example, values of the electric currents in current-carrying magnetic loops were detected with seismological methods to be about $6\times 10^{10}$--$1.4\times 10^{12}$\,A \citep{1998A&A...337..887Z} and $3\times 10^{10}$--$10^{11}$\,A \citep{2013AstL...39..650Z}. The currents in eruptive prominences and pre-eruptive filaments were found to have typical values of $10^{11}$--$10^{12}$\,A \citep{1994ChA&A..18..212W} and $4\times 10^{12}$\,A \citep[estimated from Fig.~7\,(c) of][taking the mean value of the parallel current density to be 0.05\,A\,m$^{-2}$ and the flux rope diameter $\sim 10$~Mm]{2010ApJ...715.1566C}. {For estimations, we consider a quiescent (i.e. non-eruptive) prominence of the mass density $\sim 10^{-10}\,\mathrm{kg\,m^{-3}}$, and diameter $\sim 10$~Mm, oscillating transversally with the amplitude in the range of 1--$5\ \mathrm{km\,s^{-1}}$  and 20--$100\,\mathrm{km\,s^{-1}}$, which correspond to the weakly and highly nonlinear regimes of oscillations, respectively \citep[see e.g.][]{Tripathi2009}. For such a prominence, taking the values of the normalised initial horizontal speed $\dot{x}$, used for solutions shown in Figs.~\ref{coupfig} ($\dot{x}=0.01$) and \ref{num_sol} ($\dot{x}=0.26$), the electric current in the prominence would correspond to about $1.5\times 10^{10}$--$10^{11}$\,A. These values of the prominence current are by an order of magnitude consistent with the results mentioned above, thus justifying the practical interest of both Sect.~\ref{weak} and Sect.~\ref{large} and their attribution to the weakly nonlinear and fully nonlinear regimes of oscillations, respectively.}
Furthermore, taking the prominence current $i=2.1\times 10^{10}$\,A from the detected range, additionally assuming the prominence height $h$ above the solar surface to be 26~Mm \citep{Parenti2014}, and fixing the other parameters of the model as $h/d=0.4$ and $I/i=0.5$ (providing the initial equilibrium of the prominence, determined by Eq.~(\ref{equilibrium}), to exist), the horizontal and vertical nonlinear oscillation periods would have the approximate values of 86 and 75~min, respectively. These estimations are also consistent with observations \citep[see e.g.][]{2011A&A...533A..96B}.}

Unlike the linear case considered in KNN16, {finite} amplitude horizontal and vertical oscillations are found to be coupled with each other. In a weakly nonlinear case the mode coupling is governed by set (\ref{Fx2})--(\ref{Fz2}). It represents an asymmetric nature of the mode coupling mechanism, i.e. the horizontal displacement is always able to generate the vertical displacement (see panels (a) and (d) in Fig.~\ref{coupfig}), while a pure vertical mode is fully uncoupled with the horizontal one. Such asymmetry in the coupling mechanism can be attributed to a vertical asymmetry of the initial equilibrium of the model (see Fig.~1 in KNN16). The efficiency of the coupling between the horizontal and vertical modes increases with the oscillation amplitude. In the case of oblique perturbations of the prominence, the mode coupling was detected to be more efficient for smaller angles between the direction of the initial perturbation and the horizontal axis, and is asymptotically degenerated when the prominence is perturbed almost perpendicular to the horizontal axis (see Fig.~\ref{coup_effic_fig}). For the case shown in Fig.~\ref{coup_effic_fig} with $h=0.5d$ and $I=0.5i$, the ratio of the maximum vertical and horizontal finite amplitude displacements was found to be of about 0.5--0.7, even when the initial attack angles are small (approximately up to 25 degrees with respect to the horizontal axis). The latter shows that the direction of the initial driver plays an important role in the initiation of the filament transverse oscillations. Due to strong mode coupling both vertically and horizontally polarised finite amplitude displacements can be expected to be simultaneously detectable in observations, even if the initial perturbation, for example a global coronal shock wave, is directed almost horizontally \citep[e.g.][]{2008ApJ...676L..89B, 2014ApJ...795..130S}. We would like to point out, that in addition to the mechanism based on the operation of the Kelvin--Helmholtz instability during the prominence evolution \citep[e.g.][]{Terradas2016}, our model suggests an alternative mechanism for the excitation of the filament displacements in the direction perpendicular to the direction of the initial driver by the nonlinear mode coupling.

Spatial structure and temporal evolution of transverse oscillations of the prominence in a weakly nonlinear case are described by the general analytical solution of set (\ref{Fx2})--(\ref{Fz2}), given by Eqs.~(\ref{solx})--(\ref{solz}). For a special case when the frequency of the vertical mode is twice the horizontal mode frequency, $\omega_2=2\omega_1$, solutions (\ref{solx})--(\ref{solz}) imply a nonlinear resonance. The resonant condition written through the physical parameters of the initial equilibrium of the prominence is given by Eq.~(\ref{res}) and illustrated in Fig.~\ref{regions}. Prominence oscillatory dynamics in both resonant and non-resonant cases is shown in Fig.~\ref{coupfig}. Its space trajectories exhibit a Lissajous-like behaviour, with a limit cycle of a symmetric hourglass shape (Fig.~\ref{z0fig} and panel (c) in Fig.~\ref{coupfig}), appearing in the resonant case and determined analytically by Eq.~(\ref{resz0}). Such a non-trivial polarisation of transverse oscillations is caused by the nonlinear coupling between vertical and horizontal displacements, described above, and potentially may be detected in observations \citep[][]{Hershaw2011, 2015RAA....15.1713P}.

Analysis of the fully nonlinear equations of motion (\ref{Fx})--(\ref{Fz}) allowed us to perform a comprehensive study of the prominence transverse oscillations of arbitrary amplitudes and assess the applicability of the approximate solutions. More specifically, the set of equations (\ref{Fx})--(\ref{Fz}) was found to be of a Hamiltonian form with the potential energy of the prominence, obtained in the exact analytical form in Eq.~(\ref{U}). In the range of parameters $h<d$ and $I<i$ (region I in Fig.~\ref{regions}), the potential energy (\ref{U}) was revealed to have a dip of a finite depth (panel (a) in Fig.~\ref{potentials}), corresponding to a so-called metastable state of the prominence. It is characterised by a critical value of the prominence potential energy, below which the prominence is always stable and experiences oscillations within the potential dip, and, in contrast, may escape the dip and become unstable in the horizontal direction, when its energy exceeds this threshold value. In other words, this equilibrium is stable to small amplitude oscillations, while becomes unstable when the amplitude exceeds a certain threshold.
In particular, in this regime the prominence may experience several oscillation cycles of varying polarisation, and then become unstable (see the right panel of Fig.~\ref{num_sol}). A similar behaviour {of an erupting filament} was observed by \citet{2006A&A...449L..17I}. {However, we should note here that the discussed model does not describe an eruption mechanism. It only addresses an initial loss of the prominence equilibrium, which may potentially lead to its eruption.}
Similarly to a weakly nonlinear case (Fig.~\ref{coupfig}), fully nonlinear oscillatory trajectories also have a Lissajous-like shape, which is worth searching for in the complex dynamics of oscillating prominences, detected in observations \citep[e.g.][]{2008ApJ...685..629G, 2017ApJ...836..178T}.
The maximum vertical and horizontal oscillation amplitudes, as well as the critical space contour, corresponding to that critical value of the prominence potential energy, are derived in Eqs.~(\ref{xm})--(\ref{zm}) and illustrated in Fig.~\ref{maxamp}. For a broad range of the intrinsic physical parameters of the model, $h$, $d$, $i$, $I$, and $\mathcal{R}$, determining the initial equilibrium of the prominence, the values of the maximum vertical and horizontal amplitudes were found to be close to each other by an order of magnitude, and comparable with typical geometrical sizes of the system, $h$ and $d$. In the limiting case of large distances between the external photospheric currents, $d$, when the magnetic dip is sufficiently suppressed, the maximum horizontal amplitude, $x_m$ is of about $d$, which is consistent with the Kuperus--Raadu model \citep{KR1974}.

Typical periods of horizontal and vertical oscillations as a function of the oscillation amplitude and the prominence equilibrium parameters $h$, $d$, $i$, $I$, and $\mathcal{R}$ were determined analytically in Eqs.~(\ref{nperiodx}) and (\ref{nperiodz}), respectively. The horizontal oscillation period (panels (a) and (b) in Fig.~\ref{periods}) was found to increase with the oscillation amplitude and with the height of the filament above the photosphere, which is consistent with the recent observational results \citep{Hershaw2011,2013ApJ...779L..16H}. In turn, the vertical oscillation period (panels (c) and (d) in Fig.~\ref{periods}) appears to increase with the oscillation amplitude for lower values of the ratios $h/d$ and $I/i$, and decreases for higher values of these two parameters. In the limiting cases when $h\ll d$ and $i\approx I$, horizontal oscillations were found to be nearly isochronous, i.e. the oscillation period weakly depends on the oscillation amplitude. Similarly, the approximate isochronous nature of the vertically polarised mode is detected for $h/d\approx 0.5$ and $I/i\approx 0.6$. Hence, in these special cases the analytical dependences of oscillation periods upon the intrinsic parameters of the magnetic system, derived for linear oscillations in KNN16, can be used with a good certainty for observed transverse oscillations of an arbitrary amplitude.
Another interesting feature clearly shown by Fig.~\ref{periods} is that the dependences of the horizontal period upon the amplitude for all shown examples have positive second derivatives (see panels (a) and (b)), while the corresponding dependences of the vertical period (panels (c) and (d)) are seen to have negative second derivatives. The latter fact could be straightforwardly used to distinguish between polarisations of observed large amplitude prominence oscillations as this quantity is rarely detectable without spectroscopic instruments. For example, \citet{2013ApJ...779L..16H} performed a statistical study of transverse oscillations in prominence threads. They revealed the dependence of the oscillation period, $P$ upon the oscillation amplitude, $A$ {to be in the power law form $P\propto A^{1.35}$, implying its second derivative, $0.4725\,A^{-0.65}$, is always positive.}
According to our analysis, {the positive sign of this derivative indicates that} the considered transverse oscillations are of the horizontal polarisation {(cf. Fig.~\ref{periods})}, that agrees with the results of \citet{2013ApJ...779L..16H}, where the oscillations are thought to be driven by horizontal photospheric motions.

We need to mention that the developed model should be considered as a simple one. It clearly misses a number of important physical phenomena connected with thermodynamical, partial ionisation, and {dissipative} effects. They can affect, in particular, time evolution of the oscillations, leading to their damping or amplification. Another potentially important effect is connected with the nonuniformity of the physical parameters across the plane of the model, for example the curvature of the magnetic rope \citep{1994ApJ...423..854C, 2008AnGeo..26.3089V}. {The line-tying boundary conditions for the guiding magnetic field would add an additional force, affecting the estimated oscillation periods and the stability conditions. It makes our model acceptable only when the height $h$ of the magnetic rope axis above the solar surface is much smaller than the distance between the footpoints of the guiding field, and an eventual axial component of the guiding field is therefore strictly aligned with the \textit{y}-axis.} {Likewise, the effects of an inverse magnetic polarity of the prominence could also be investigated in terms of this model, which can alter the values of oscillation periods and threshold amplitudes.} These and other phenomena may be taken into account in the further development of the model. Nevertheless, we believe that the model sufficiently advances our understanding of prominence oscillations, attracts attention to the important observables, such as the oscillation polarisation and finite amplitude, demonstrates the existence of metastable equilibria, and provides a foundation for seismological estimation of the value of the macroscopic current in coronal magnetic ropes.

\section*{Acknowledgements}
\noindent DYK acknowledges the support of the STFC consolidated grant ST/L000733/1.
VMN acknowledges the support of the Russian Science Foundation under grant 16-12-10448. 

\appendix
\section{Solution of set (\ref{eqA1})--(\ref{eqA2})}\label{app1}
From Eq.~(\ref{eqA1}) the function $A_2(\tau)$ can be expressed as
\begin{equation}\label{A2(A1)}
A_2=\frac{2\omega_2}{\beta A_1}\frac{d\,A_1}{d\,\tau}.
\end{equation}
Using (\ref{A2(A1)}), Eq.~(\ref{eqA2}) reduces to the following second order ordinary differential equation
\begin{equation}\label{ODE_A1}
\frac{1}{A_1}\frac{d^2A_1}{d\,\tau^2}-\frac{1}{A_1^2}\left(\frac{d\,A_1}{d\,\tau}\right)^2+\lambda A_1^2=0,
\end{equation}
with $\lambda={\beta\delta}/{8\omega_2^2}$. Then writing $d\,A_1/d\,\tau=P(A_1)$, and hence $d^2A_1/d\,\tau^2=P(d\,P/d\,A_1)$, Eq.~(\ref{ODE_A1}) takes the form
\begin{equation}\label{ODE_P(A1)}
\frac{1}{A_1}\frac{d}{d\,A_1}\left(\frac{P^2}{2}\right)-\frac{P^2}{A_1^2}=-\lambda A_1^2.
\end{equation}
With the use of a new variable $s=A_1^2$, so that $d/d\,A_1=2A_1(d/d\,s)$, Eq.~(\ref{ODE_P(A1)}) can be re-written as
\begin{equation}\label{ODE_P(s)}
\frac{d\,P^2}{d\,s}-\frac{P^2}{s}=-\lambda s.
\end{equation}
Now expressing the function $P$ through a new unknown function $q(s)$ as $P^2=sq(s)$, Eq.~(\ref{ODE_P(s)}) goes to
\begin{equation}\label{ODE_q(s)}
\frac{d\,q}{d\,s}=-\lambda,
\end{equation}
which can be integrated once and has the solution
\begin{equation}\label{q(s)}
q=q_0-\lambda s,
\end{equation}
with $q_0$ being a constant determined from the initial conditions $A_1(0)=A_0$ and $\dot{A}_1(0)=0$, as $q_0=\lambda A_0^2$.

Now recalling that $P=d\,A_1/d\,\tau$, $P^2=sq(s)$, and $s=A_1^2$, we obtain the following equation
\begin{equation}\label{ODE_A1_2}
\left(\frac{d\,A_1}{d\,\tau}\right)^2=A_1^2(q_0-\lambda A_1^2),
\end{equation}
whose integral has the form
\begin{equation}\label{tau(A1)}
\tau=\int_{A_0}^{A_1}\frac{d\,A_1}{A_1\sqrt{q_0-\lambda A_1^2}}=\frac{1}{A_0(\lambda)^{1/2}}\sech^{-1}\left(\frac{A_1}{A_0}\right).
\end{equation}
Using (\ref{tau(A1)}), we are able to write the explicit solution $A_1(\tau)$ as
\begin{equation}\label{A1(tau)}
A_1=A_0\sech[A_0(\lambda)^{1/2}\tau].
\end{equation}
Substitution of (\ref{A1(tau)}) into (\ref{A2(A1)}) gives the explicit form of the dependence $A_2(\tau)$,
\begin{equation}\label{A2(tau)}
A_2=-A_0\left(\frac{\delta}{2\beta}\right)^{1/2}\tanh[A_0(\lambda)^{1/2}\tau].
\end{equation}

Having obtained the explicit solutions for $A_1(\tau)$ and $A_2(\tau)$, we use them in the lowest order harmonic solutions given in (\ref{harmonicx2})--(\ref{harmonicz2}) to obtain
\begin{equation}\label{x0(t,tau)}
x_0=A_0\sech[A_0(\lambda)^{1/2}\tau]\sin(\omega_1t),
\end{equation}
\begin{equation}\label{z0(t,tau)}
z_0=-A_0\left(\frac{\delta}{2\beta}\right)^{1/2}\tanh[A_0(\lambda)^{1/2}\tau]\sin(\omega_2t).
\end{equation}
Finally, using the resonant condition $\omega_2=2\omega_1$ one can obtain an explicit relation between the vertical and horizontal coordinates, $z_0$ and $x_0$, describing the prominence space dynamics in the special resonant case:
\begin{equation}
z_0^2=\frac{2\delta}{\beta}x_0^2\sinh^2[A_0(\lambda)^{1/2}\tau]\left\{1-\frac{x_0^2}{A_0^2}\cosh^2[A_0(\lambda)^{1/2}\tau]\right\}.
\end{equation}


\end{document}